\documentclass[twocolumn,10pt,aps,prd,floatfix,nofootinbib]{revtex4}

\usepackage {amsmath}
\usepackage{amssymb}
\usepackage{amsfonts}
\usepackage{graphicx}
\usepackage{array}
\usepackage{booktabs}
\usepackage{psfrag}

\newcommand{\openinflation}{1995ApJ...455..412Y,
  1996PhRvD..54.5031Y,1997PhRvD..55.4603G, 
  1999PhRvD..59l3522L, 1999PhRvD..60h3501G}

\begin{document}
\title{Anisotropic Kantowski-Sachs Universe from Gravitational
  Tunneling and its Observational Signatures}
\author{Julian Adamek}
\email{jadamek@physik.uni-wuerzburg.de}
\affiliation{Institut f\"ur Theoretische Physik und Astrophysik, 
         \\Julius-Maximilians-Universit\"at W\"urzburg,\\ Am Hubland, 97074 W\"urzburg, Germany}
\author{David Campo}
\email[]{dcampo@astro.physik.uni-goettingen.de}
\author{Jens C.\ Niemeyer}
\email[]{niemeyer@astro.physik.uni-goettingen.de}
\affiliation{Institut f\"ur Astrophysik, 
         \\Georg-August-Universit\"at G\"ottingen,\\ Friedrich-Hund-Platz 1, 37077 G\"ottingen, Germany}

\begin{abstract}
In a landscape of compactifications with different numbers of
macroscopic dimensions, it is possible that our universe has nucleated
from a vacuum where some of our four large dimensions were compact
while other, now compact, directions were macroscopic. From our
perspective, this shapeshifting can be perceived as an anisotropic
background spacetime.
As an example, we present a model where our universe emerged
from a tunneling event which involves the decompactification 
of two dimensions compactified on the two-sphere.
In this case, our universe is of the Kantowski-Sachs type and therefore  
homogeneous and anisotropic.
We study the deviations from statistical isotropy of the Cosmic Microwave
Background induced by the anisotropic curvature, with particular attention to the anomalies. 
The model predicts a quadrupolar power asymmetry 
with the same sign and acoustic oscillations as found by WMAP.
The amplitude of the effect is however too small given the current 
estimated bound on anisotropic curvature derived from the quadrupole.
\end{abstract}
\maketitle

\def\dd{{\mathrm{d}}}


\section{Introduction}

If inflation lasted sufficiently long, all curvature scales
imprinted on our universe by pre-inflationary physics are pushed to
undetectable distances beyond our current horizon. If, on the other
hand,  
inflation ended soon after the required amount of accelerated expansion to allow 
a later epoch of rich structure formation in our local universe,  
this cosmic amnesia may have been only partial. The largest cosmological
scales observable today would then potentially show traces of the initial conditions for
our inflating universe, including curvature \cite{2006JHEP...03..039F,
\openinflation}, anisotropies
\cite{GCP,PPU1,PPU2}, or
nontrivial topology \cite{LachiezeRey:1995kj}. This is the setting in 
which the scenario we propose can have observational
consequences. 
While it has often been stated that fine-tuning the
amount of inflation to this extent is unnatural, the landscape
paradigm combined with the difficulty to find long-lasting
inflationary solutions in string theory provide
sufficiently strong counterarguments to take this
possibility
seriously \cite{DeSimone:2009dq}. Like the collision of bubbles in scenarios of 
false vacuum inflation (see \cite{Aguirre:2009ug} for a status
report), it offers a remote chance to probe the landscape of string
theory by observations.

What are plausible initial conditions for our local inflationary patch
in the landscape? 
Compactification in string theory is often treated kinematically, 
as part of the construction of the effective four dimensional theory.
With four macroscopic plus 
a number of microscopic compact dimensions fixed once and for all, 
the transition between
metastable vacua can be described by the spontaneous nucleation of
bubbles with open homogeneous and isotropic spatial
sections \cite{CDL,Aguirre:2005nt}. The observable
consequences of false vacuum bubble nucleation 
followed by a brief period of slow-roll inflation
have been studied extensively in the context of open inflation
\cite{\openinflation} and are well understood by now.

From the dynamical perspective, however, compactification
becomes a problem of string cosmology, for the compactified dimensions
can spontaneously
open up and become large \cite{Giddings}.
This substantially widens the parameter space for initial
conditions of our local universe,  
since there are now alternative channels to populate the landscape 
which can be viewed as
transitions between vacua with differing numbers of macroscopic
dimensions. This was first explicitly spelled
out by Carroll, Johnson, and Randall \cite{CJR} in the context of
dynamical compactification from a higher dimensional spacetime to our
effectively four-dimensional one. 
The opposite process, dynamical decompactification, was studied in
\cite{BlancoPillado:2009di}, and in \cite{GiddingsMyers} in a different context. 

Although these previous studies were all concerned with the 
(de-)compactification of higher dimensions, there is no reason 
to exclude that the three macroscopic dimensions
of our present universe may themselves be the result of such a process
\cite{BrandyVafa}. This is the starting point of our work\footnote{Shortly before
  completion of our work, two articles (\cite{GHR} and \cite{BPS})
  were posted which have a significant amount of overlap with ours. We
will comment on the similarities and differences at various places in
the main text.}.
It is intuitively clear that decompactification allows the existence 
of a preferred direction in the sky 
if only one or two directions are compact,
and therefore gives rise to anisotropic cosmologies.
We specifically consider the case where two of our macroscopic
dimensions are compact. Before inflation, these 
dimensions were microscopic, leaving one macroscopic direction which
may still play a preferred role for cosmological observations today if
inflation was short. 
As a concrete example, we present a four dimensional model with two
dimensions compactified on a sphere 
by the flux of an Abelian gauge field and a cosmological constant.
The solutions of the Euclidean Einstein-Maxwell equations with spherical symmetry 
are well known. They describe pair creation of charged black holes 
in de Sitter space \cite{BoussoHawking1,BoussoHawking2,MannRoss}.
We note that the causal patch beyond the cosmological 
horizon in the Lorentzian spacetime is an anisotropic inflating Kantowski-Sachs (KS) universe. 

This initial state can be placed in the broader context of the
landscape in different ways. It can be viewed as an intermediate phase
in a progressive opening up of initially compact dimensions as
described in \cite{GiddingsMyers}, or as the
temporarily final step in a sequence of transitions starting from a 
higher dimensional geometry that triggered the 
decompactification of two previously compact dimensions. We comment
briefly on these scenarios in Sec.\ \ref{ref:motivations} but we will
leave the discussion of further implications for future work. 

The remainder of this article is structured as follows. In Sec.\
\ref{sec:model}, the model is introduced and its 
connection to
black hole pair production is discussed. Sec.\
\ref{sec:perturbations} presents a preliminary exploration of
Cosmic Microwave Background (CMB) signatures 
in our model, which are due to the anisotropy of KS spacetime.
Owing to
the technical complexity of a full analysis, we only consider the
perturbations of a test scalar field and outline the qualitative
modifications of CMB temperature anisotropies. 
With these results we try to assess some of the observed CMB anomalies
which seem to 
indicate a violation of statistical isotropy in our universe.
We conclude and discuss some directions for further research along these
lines in Sec.\ \ref{sec:conclusions}.


\section{The shapeshifting universe}
\label{ref:motivations}

\subsection{Rearrangement of macroscopic dimensions}

The hypothesis of a string landscape \cite{Susskind,BP} leads to the 
following scenario. The effective potential of the 
moduli (scalar fields describing the 
size and shapes of compact dimensions) possesses local minima,
created from the competing effects of a positive cosmological constant, 
the curvature of the compact spaces, fluxes, wrapped branes, etc., but
also presents flat directions. 
The compactified configurations, which are only metastable 
as a consequence of gravitational tunneling, may thus open up 
\cite{Giddings,GiddingsMyers}. Consequently, 
in the general case the four-dimensional macroscopic spacetime has
one, two, or three compact 
spatial dimensions, and is anisotropic (one exceptional, and
interesting case, is the compactification on a flat torus as in
\cite{OVV}, see also \cite{McInnes}). 

More precisely, we 
consider
the mechanism where $q$ dimensions are compactified on 
a $q$-sphere
whose radius is stabilized using the
flux of an appropriate $q$-form field \cite{Grana:2005jc,Krishnan:2005su}.
Many such 
gauge fields are available in string theories, 
each one coming with its own gauge coupling. 
One therefore obtains
an entire landscape of possible flux
compactifications \cite{BP}. The vacua in this landscape 
(the minima of the effective potential) differ in the
number of large dimensions and in the effective value of their vacuum
energy, which is determined by the quantum numbers of the flux
fields. 

These vacua are rendered metastable by the possibility to tunnel
through the barriers in the effective potential of the moduli. 
In such a process, a bubble is formed containing a new vacuum 
configuration, which differs
from the parent vacuum in the values of the fluxes and, possibly, in
the number of compactified dimensions. 
Elementary transitions where compact
dimensions destabilize and start to open up (decompactification) and
the inverse process (compactification) have been investigated by
\cite{GiddingsMyers} and \cite{CJR}, respectively.
It has been recognized that both processes can be described with a generic
type of instanton, and that the two different interpretations follow
by exchanging the labels for ``parent'' and ``daughter'' vacuum.
More generally, however, a vacuum transition may 
re-arrange the configuration of compact
directions in an arbitrary way, such that it is possible that some
dimensions compactify while others decompactify at the same
time\footnote{This has already been speculated in \cite{BSV}.}.
We are thus confronted with the interesting possibility
of a shapeshifting universe:
some of our macroscopic dimensions 
may have been destabilized
in the final vacuum transition which spawned our present universe
and have been opening up ever since.

\begin{figure}[t]
\includegraphics[width=85mm]{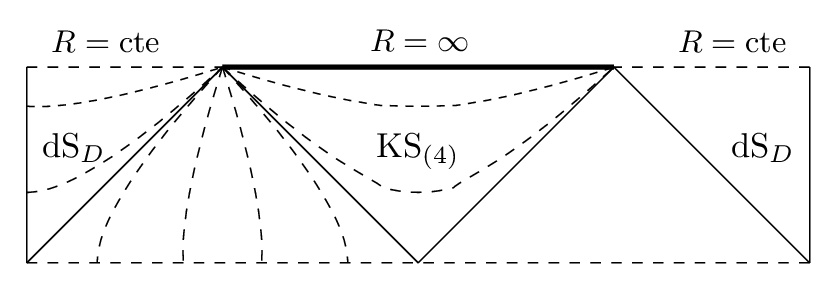}
\caption{\label{fig:shape-shift} \small 
Spacetime diagram of a generic tunneling process giving rise to a
four-dimensional Kantowski-Sachs universe. 
The effective $D$-dimensional parent vacuum is marked as $\mathrm{dS}_D$
(left and right edge of the diagram should be identified). 
In this region, the two compact directions of our
macroscopic universe approach a constant microscopic radius, as they
are stabilized on $\mathrm{S}_2$ by a magnetic flux. In the 
Kantowski-Sachs spacetime, marked as $\mathrm{KS}_{\left(4\right)}$, 
these directions have destabilized and become large.
Vice versa, there are $D-2$ large dimensions in the parent 
vacuum which approach a compact microscopic configuration
in our universe. The ``static'' regions interpolate between 
the parent vacuum and our bubble. The interpolation smoothly
re-arranges all the moduli fields to their new vacuum configuration. 
The case where the parent vacuum shares two large
dimensions with our universe has recently been studied in \cite{BPS}. 
In this case, the $\mathrm{KS}_{\left(4\right)}$ region 
with topology $\mathbb{R}\times \mathrm{S}_2$ 
is replaced by a Bianchi III spacetime, which has spatial 
topology $\mathrm{H}_2 \times \mathrm{S}_1$. The familiar
result of an open FRW universe in place of $\mathrm{KS}_{\left(4\right)}$ 
is obtained if all of our macroscopic dimensions
are shared by our parent vacuum. }
\end{figure}

The outcome of such re-arrangements is that the macroscopic space dimensions
have a nontrivial topology, for instance $\mathbb{R}\times \mathrm{S}_2$
corresponding to a KS spacetime, 
or $\mathrm{H}_2 \times \mathrm{S}_1$ corresponding to Bianchi III.
Schematically, this history can be described as
$$
\mathrm{dS}_D \times \mathrm{S}_2 \times \mathcal{M}_d \longrightarrow
\mathrm{KS}_{\left(4\right)} \times \mathcal{M}'_{d + D - 2}~, 
$$
where $\mathrm{dS}_D$ is our effective $D$-dimensional parent vacuum,
$\mathrm{KS}_{\left(4\right)}$ is the Kantowski-Sachs spacetime describing our macroscopic
universe, and $\mathcal{M}_d$, $\mathcal{M}'_{d + D - 2}$ are
additional compactification manifolds. The corresponding
Penrose-Carter diagram is shown in Fig.~\ref{fig:shape-shift}. 
Owing to the reconfiguration of the
microscopic dimensions, the effective vacuum energy is changed with
respect to the parent vacuum and may lie in the anthropic window if
the usual conditions for the smallness of incremental changes apply
\cite{BP,CJR}.

\subsection{Description of the model}

The dynamics of 
tunneling processes involving several interacting fields (the moduli) 
can be very complicated (e.g., \cite{Johnson:2008kc}). 
Here, however, we are mainly interested in the geometrical properties
of the corresponding instanton. Furthermore, we will treat the microscopic
dimensions in our present universe as mere spectators, and therefore
we can effectively work in a four-dimensional description.
The tunneling event describing the decompactification of two dimensions
wrapped onto a two-sphere is then
completely equivalent to the pair creation of charged black holes. It
was already pointed out in \cite{CJR} that the mechanism of dynamical
compactification, using magnetic fluxes, is in some sense a
generalization of the idea of black hole pair creation. The
connection between these ways of thinking was further elucidated in
\cite{BSV}, where different types of compactifications from a
six-dimensional de~Sitter vacuum have been studied.

Ignoring the additional dimensions in our description comes at the cost 
of not being able to represent the parent vacuum correctly, nor making quantitative
statements about the tunneling rates. However, our simplified picture should
accurately capture the essential geometrical properties of the daughter vacuum.
In particular, within our scenario
where two directions are wrapped onto a sphere, 
the tunneling process gives rise to
a Kantowski-Sachs universe of topology $\mathbb{R} \times \mathrm{S}_2$.

If the scales comparable to the size of the compact directions have
recently entered our horizon, we may gather information about the
topology of our universe by examining, e.g., the temperature
correlations in the CMB. Inflation is still a necessary ingredient in
this class of models in order to dilute the spatial curvature, produce
a phenomenologically viable perturbation spectrum, and heat up the
universe. 
However, any detection of the CMB
signatures that we discuss in Sec.\ IV implies that inflation ended
before they were redshifted too far beyond the horizon. 
Whether or not anthropic pressure makes this
parameter range likely is an interesting debate (e.g., \cite{DeSimone:2009dq}) to
which we have nothing new to add. For the purposes of this work, we
simply accept it as a plausible possibility.

As the authors of \cite{CJR} have shown, it is possible to accommodate
a period of slow-roll inflation following the transdimensional
tunneling. In their example, this is achieved by coupling an
inflaton field to the curvature and flux. It was demonstrated that the
tunneling process can indeed trigger slow-roll inflation by setting
free the inflaton, which was previously trapped in a minimum by the
configuration of the parent vacuum.  In this case, the initial conditions for
inflation are 
governed by the tunneling process. In order to simplify
our analysis, we do not consider a realistic model for slow-roll
inflation and absorb the inflaton potential energy into the effective
cosmological constant. Thus, in our model $\Lambda$ is the sum of the 
inflaton energy density and of the observed value of the cosmological constant.


\section{Kantowski-Sachs universe from gravitational tunneling in 
$4$D Einstein-Maxwell theory}
\label{sec:model}

We now introduce our cosmological model and 
outline the links with the pair creation of charged black holes 
during inflation. 
Most of the content of this section is 
well-known
and was originally 
presented in the context of black hole pair creation (see, e.g.,
\cite{BoussoHawking1,BoussoHawking2,MannRoss} and references
therein). In our notation and terminology, we closely follow
\cite{MannRoss}.

We present our model using a unified picture  
where pair creation of charged black holes can equivalently be 
understood as Coleman-De~Luccia tunneling between two different vacua of
a lower dimensional effective theory. We will therefore use both
interpretations interchangeably, depending on which of them is more
useful in a particular situation. 

As a starting point we take the action of 4D Einstein-Maxwell theory,
\begin{equation}
\mathcal{S} = \frac{1}{16 \pi} \int\!\dd^4 x \sqrt{-g} \Bigl[\mathcal{R}[g_{\mu\nu}] - 2 \Lambda - F_{\mu\nu} F^{\mu\nu}\Bigr]~,
\end{equation}
where a cosmological constant $\Lambda > 0$ is included in order 
to model inflation in a simple way.
$\mathcal{R}[g_{\mu\nu}]$ is the Ricci scalar of the metric tensor $g_{\mu\nu}$, 
$g$ is the metric determinant, and $F_{\mu\nu}$ is the field strength 
tensor of an Abelian gauge field.

We look for solutions with 
two of the three spatial dimensions compactified on a sphere of radius $R$. 
The line element therefore takes the form
\begin{equation}
\dd s^2 = \gamma_{ab}(\mathbf{x}) \dd x^a \dd x^b + R^2(\mathbf{x}) \left[ \dd \theta^2 + \sin^2\theta \dd \phi^2\right]~,
\end{equation}
where $\gamma_{ab}$ is the metric of a 1+1-dimensional Lorentzian manifold with coordinates
$\mathbf{x}$, i.e., the Latin indices take values $0$ and $1$ only.

The gauge field should also respect the symmetries of our metric ansatz. 
The magnetic solutions of Maxwell's equations are therefore 
proportional to the volume form of $\mathrm{S}_2$,
\begin{equation}
\mathbf{F} = Q \sin\theta \dd \theta \wedge \dd \phi~,
\end{equation}
where $Q$ is the magnetic charge.

Our next step will be the dimensional reduction of the 
theory by integration over the coordinates
$\theta$, $\phi$ of $\mathrm{S}_2$. Decomposing the full Ricci scalar $\mathcal{R}[g_{\mu\nu}]$
into contributions from the $\mathrm{S}_2$-curvature and from the Ricci scalar of $\gamma_{ab}$, hereafter
denoted as $\mathcal{R}[\gamma_{ab}]$, the action can be rewritten as 
\begin{multline}
\mathcal{S} = \frac{1}{4} \int \dd^2 x \sqrt{-\gamma} \Bigl[ R^2 \mathcal{R}[\gamma_{ab}] + 2 \gamma^{ab} \nabla_a R \nabla_b R \Bigr. \\ \Bigl. + 2 - 2 \Lambda R^2 - \frac{2 Q^2}{R^2}\Bigr]~,
\end{multline}
after an integration by parts has removed second derivatives on $R$. Variation with respect to $R$ and $\gamma_{ab}$ yields
two coupled second order equations,
\begin{equation}
\label{EOM}
\frac{1}{2} \mathcal{R}[\gamma_{ab}] R - \square R - \Lambda R + \frac{Q^2}{R^3} = 0~,
\end{equation}
and
\begin{multline}
\label{EEQ}
2 R \nabla_a \nabla_b R - 2 \gamma_{ab} R \square R - \gamma_{ab} \gamma^{cd} \nabla_c R \nabla_d R \\+ \gamma_{ab} \Bigl( 1 - \Lambda R^2 - \frac{Q^2}{R^2} \Bigr) = 0~.
\end{multline}

Following the usual procedure, we now solve the corresponding Euclidean equations 
with the boundary conditions appropriate to describe a tunneling Riemannian geometry. 
In section \ref{sec:cosmology}, we consider the nucleated $4$D Lorentzian geometries
which contain a KS spacetime.
These solutions describe a universe with two compact 
spatial dimensions which are destabilized and start to grow. 

\subsection{Euclidean geometries}
\label{sec:instantons}

A formal analytic continuation to imaginary time
takes the action to its Euclidean counterpart. 
The Euclidean equations remain formally identical
to the Lorentzian ones (\ref{EOM}) and (\ref{EEQ}), with the important
difference that the metric $\gamma_{ab}$ now has Euclidean signature $(++)$.

We consider all the solutions of the $2$D Euclidean equations which have 
$O(2)$-symmetry, meaning that
all quantities only depend on the Euclidean distance $\chi$ from the symmetry point:
\begin{equation}
\begin{array}{r c l}
\gamma_{ab} \dd x^a \dd x^b &=& \dd \chi^2 + \rho^2(\chi) \dd \varphi^2~, \\
R(\chi, \varphi) &=& R(\chi)~.
\end{array} \quad \text{(Euclidean)}
\end{equation}
Here, the coordinate $\varphi$ is an angular coordinate with period $2 \pi$.
We will relate these solutions to the instantons describing pair creation of 
charged black holes in de~Sitter space.

At the symmetry point $\chi = 0$ the Euclidean scale factor $\rho$ is zero. The zeros of $\rho$ are sometimes
called ``poles'' since they may represent the origin of a polar coordinate system. In this paper, we will
also call the symmetry point $\chi = 0$ the ``south pole'' of the geometry. If $\rho$ has a second zero at
finite $\chi = \chi_{\mathrm{max}}$, we will call this the ``north pole''. Regularity of the geometry at
the poles requires $\rho'(0) = -\rho'(\chi_{\mathrm{max}}) = 1$, otherwise there will be a conical singularity.

Using the $O(2)$-symmetric ansatz, eqs.~(\ref{EOM}) and (\ref{EEQ}) read, respectively,
\begin{equation}
\label{EucEOM}
\frac{\rho''}{\rho} R + R'' + \frac{\rho'}{\rho} R' + \Lambda R - \frac{Q^2}{R^3} = 0~,
\end{equation}
and
\begin{equation}
\label{EucEEQ}
R'^2 + 2 \frac{\rho'}{\rho} R' R - 1 + \Lambda R^2 + \frac{Q^2}{R^2} = 0~.
\end{equation}

These two equations can be combined to show that $\dd \ln \rho = \dd \ln R'$, which implies
$\rho \propto R'$. This result can be used to eliminate $\rho$ in eq.~(\ref{EucEEQ}). One finds
\begin{equation}
2 R R'' + R'^2 - 1 + \Lambda R^2 + \frac{Q^2}{R^2} = 0~,
\end{equation}
and its first integral
\begin{equation}
\label{ECL}
R'^2 - \frac{Q^2}{R^2} + \frac{2 M}{R} - 1 + \frac{\Lambda}{3} R^2 = 0~.
\end{equation}
These solutions are indeed the Euclidean analogs of Reissner-Nordstr\"om-de~Sitter (RNdS)
black holes, where the constant of integration $M$ is the Misner-Sharp mass.
The positive roots of the potential
\begin{equation}
V(R) \equiv - \frac{Q^2}{R^2} + \frac{2 M}{R} - 1 + \frac{\Lambda}{3} R^2
\end{equation}
correspond to the horizons of the black hole spacetime. These are, in ascending order, the inner
Cauchy horizon $R_i$, the black hole event horizon $R_b$, and the cosmological horizon $R_c$.
There is also one root at negative $R$ which has no physical significance. 

\begin{figure}[t]
\includegraphics[width=85mm]{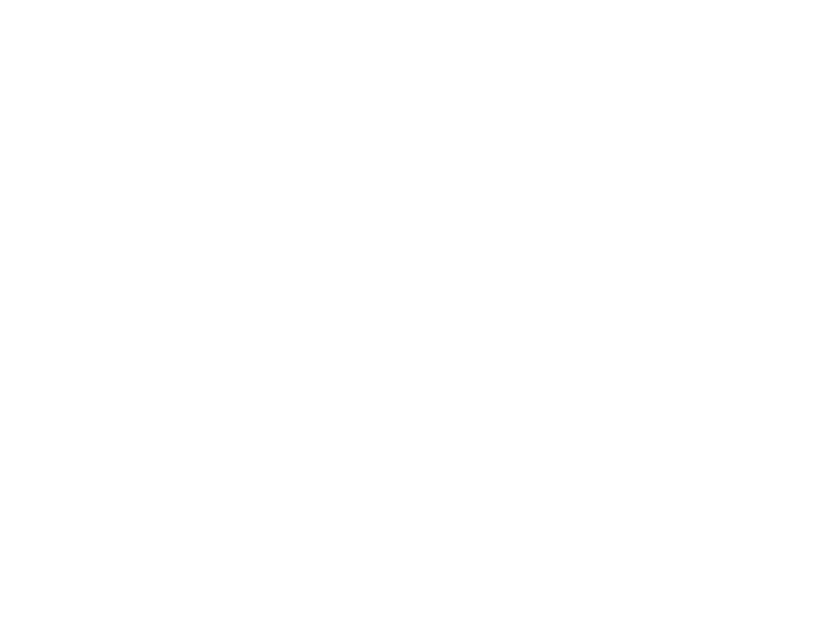}
\caption{\label{fig:potential} \small Qualitative shape of the potential $V(R)$ for three different
mass parameters $M$ at fixed $\Lambda$ and $Q$, with $Q < 3 / \sqrt{48 \Lambda}$. 
The roots of $V(R)$ correspond to the three different horizons in a
Reissner-Nordstr\"om-de~Sitter spacetime. The lower dashed line (smallest value of $M$) shows the special
case when the black hole event horizon $R_b$ and the inner Cauchy horizon $R_i$ coincide
(extremal Reissner-Nordstr\"om black hole). This potential may characterize a ``cold'' instanton as
well as the $\mathrm{H}_2 \times \mathrm{S}_2$ instanton, which has no finite action. The upper dashed line
(largest value of $M$) shows the special case when $R_b$ coincides
with the cosmological horizon $R_c$. This would be characteristic for a ``charged Nariai'' instanton.
The line in between shows the case when $V'(R_b) = -V'(R_c)$, corresponding to a ``lukewarm'' instanton.}
\end{figure}

The number of positive roots can of course be less than three. For a given $\Lambda$ and a
sufficiently small $Q$, there is a minimal and a maximal value for $M$ at which two roots
exactly coincide, such that $V(R)$ has a double root. These two situations are realized when
$R_i \equiv R_b$ and $R_b \equiv R_c$, respectively. For the intermediate range of $M$, all three
roots are distinct. This range for $M$ shrinks as $Q$ increases, and at $Q = 1 / \sqrt{4 \Lambda}$,
all three positive roots coincide for $M = 2 / \sqrt{18 \Lambda}$. For $Q$ larger than that, or for
$M$ outside the intermediate range, only one positive root exists, and no instanton of finite action can
be obtained. Let us first examine the cases where $M$ is chosen such that $V(R)$ has a double root.

If $R_b$ and $R_c$ coincide, they designate a minimum of the potential, see Fig.\ \ref{fig:potential}.
Hence, there is a solution $R = \mathrm{const.} = R_b (\equiv R_c)$. In this case
$\rho = H^{-1} \sin H \chi$, with $H^2 = V''(R_b) / 2$. 
[The easiest way to see this
is to write $R = R_b + \delta R$ and expand eq.\ (\ref{ECL}) to second order in the
perturbation $\delta R$. The functional form of $\rho$ then follows from $\rho \propto \delta R'$,
and since the amplitude of $\rho$ is entirely fixed by the boundary conditions $\rho(0) = 0$,
$\rho'(0) = 1$, it is independent of the amplitude of the
perturbation.] 
The Euclidean solution
is a sphere with radius $H^{-1}$. This particular solution is called the charged Nariai instanton,
since the analytic continuation to Lorentzian signature turns the $\mathrm{S}_2 \times \mathrm{S}_2$,
obtained in the full 4D picture, into a $\mathrm{dS}_2 \times \mathrm{S_2}$ (charged) Nariai spacetime.

If, on the other hand, $R_b$ and $R_i$ coincide, they designate a maximum of the potential.
There is again a solution $R = \mathrm{const.} = R_b (\equiv R_i)$, but this time one finds
that $\rho = \omega^{-1} \sinh \omega \chi$, with $\omega^2 = - V''(R_b) / 2$. The geometry
is now a hyperbolic plane $\mathrm{H}_2$ instead of a sphere. Since the hyperbolic plane has
infinite volume, this instanton does not have a finite action. This is
the reason why it 
is
usually not discussed in the context of black hole pair creation. The spacetime one would get
from analytic continuation of this instanton is $\mathrm{AdS}_2 \times \mathrm{S_2}$, again
in the 4D picture. We note that in the 2D picture the compactification vacuum has negative
effective vacuum energy, and hence the universe is Anti-de~Sitter. 
This seems to be a generic feature of two-dimensional vacua obtained by flux compactification.
However, this does not exclude the possibility that other compactification mechanisms may
give rise to effective two-dimensional solutions with positive vacuum energy. Then there
would also be two-dimensional de~Sitter vacua which could play the role of our parent vacuum.

It is evident from Fig.\ \ref{fig:potential} that in the case $R_b \equiv R_i$, a second
instanton should exist. It is obtained by starting from $R = R_c$ at the south pole of
the geometry. As $\chi$ increases, $R$ rolls through the potential well and approaches
$R = R_b$. However, since $R_b$ is a double root, $R$ reaches this value
only at $\chi = \infty$. 
The resulting geometry is a two-dimensional $O(2)$-symmetric space which interpolates
between a sphere with Gaussian curvature $V''(R_c) / 2$ at the south pole and a pseudosphere
with (negative) Gaussian curvature $V''(R_b) / 2$ as $\chi \rightarrow \infty$. It is called
the ``cold'' instanton, since it describes pair creation of extremal (and therefore cold) black
holes. The Euclidean action of this instanton is finite and was computed, e.g., in \cite{MannRoss}.

Another special class of solutions exists when all three horizons coincide, $R_i \equiv R_b \equiv R_c$.
Solutions of this type are called ``ultracold''.
Because they are only obtained 
on a single point in parameter space, namely $Q = 1/\sqrt{4\Lambda}$, 
we do not discuss them in more detail.

Let us finally turn to the case where there are three distinct horizons. Starting again with
$R = R_c$ at the south pole, one can see from Fig.\ \ref{fig:potential} that now $R$ reaches
the value $R_b$ at a finite Euclidean distance $\chi = \chi_\mathrm{max}$. Imposing the regularity
condition $\rho'(0) = 1$ at the south pole,
we see from eq.\ (\ref{ECL}) and $\rho \propto R'$ that
a conical singularity at the north pole can only be
avoided if $V'(R_c) \equiv -V'(R_b)$, which is only possible for $Q < 3 / \sqrt{48 \Lambda}$ and
fixes the mass to $M = Q$. The instanton obtained this way is regular and has topology
$\mathrm{S}_2$, and its action was also computed in \cite{MannRoss}. It describes pair creation
of non-extremal charged black holes and is therefore called
the ``lukewarm'' solution. 
In the terminology of \cite{CJR}, it corresponds to the ``interpolating solution''.

This completes our catalog of $O(2)$-symmetric solutions of the 2D Euclidean equations.
The lukewarm solutions are the only 
ones that we shall consider in the following, though the precise values of the parameters
$Q$ and $M$ are not important for our discussion.
We want to point out here that additional compact dimensions would not change the geometrical
properties of the instanton from the 2D point of view. The additional dimensions would add
new moduli fields which complicate the equations, but the $O(2)$-symmetry and the 2D topology
of the instanton would remain unchanged.

\subsection{Cosmology in the spacetime of pair-created charged black holes}
\label{sec:cosmology}

\begin{figure}[t]
\includegraphics[width=85mm]{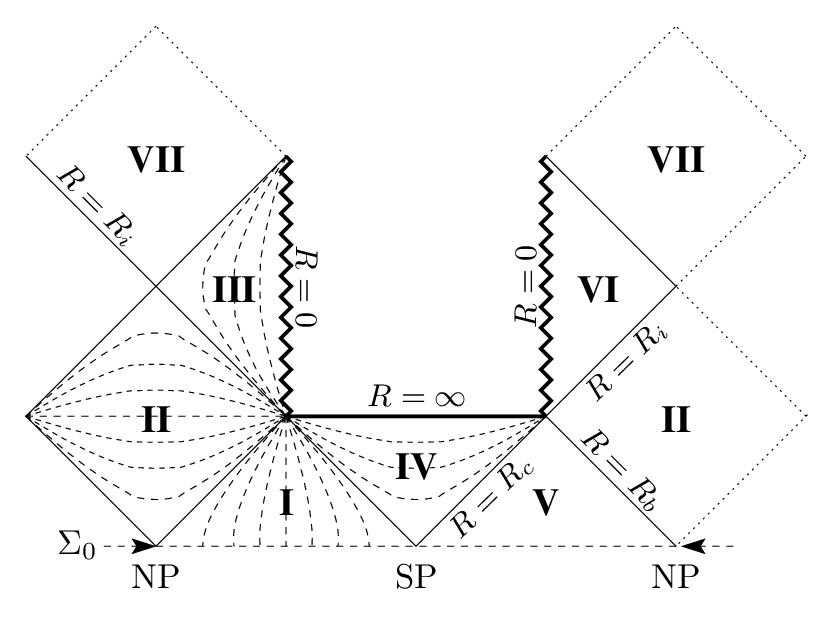}
\caption{\label{fig:pcd} \small Penrose-Carter diagram of Reissner-Nordstr\"om-de~Sitter spacetime --
left and right side (regions II and VII) should be identified. The line marked
$\Sigma_0$ shows the location of the turning point three-geometry in the diagram. It also
contains the ``north pole'' (NP) and the ``south pole'' (SP) of the instanton. Region I and V
are the static regions between the cosmological horizons and the event horizons of the two black holes.
Region II is the interior of the black holes, which make up a wormhole. It is a Kantowski-Sachs universe
evolving from a Big Bang at $R = R_b$ to a Big Crunch at $R = R_i$. Region IV lies beyond the cosmological
horizons of the black holes. It is another Kantowski-Sachs universe which has its Big Bang at $R = R_c$
and starts to inflate indefinitely due to the presence of a cosmological constant.
Regions III and VI lie beyond the inner Cauchy horizons $R = R_i$ and exist only in the mathematical solution,
as well as region VII, which would lie in another universe. In regions I-IV, curves of constant $R$ are
sketched as dashed lines. We have omitted the time-reversed copy of the diagram below $\Sigma_0$, which would
be present in a full classical solution due to time reflection symmetry. After the gravitational
tunneling event has taken place, classical evolution ``begins'' on $\Sigma_0$.}
\end{figure}

In the language of gravitational tunneling, the Euclidean geometries describe the solution
in the classically forbidden region of configuration space. A turning-point configuration,
i.e., a spacelike hypersurface which contains the initial data to solve the Cauchy
problem in the classically allowed region, is given by a maximal section of the instanton,
running perpendicular to the circles of constant $\chi$ from the south pole to the north pole
and back again. On this surface, which has topology $\mathrm{S}_1 \times \mathrm{S}_2$ in the
4D picture, the Euclidean solution connects to a classical Lorentzian spacetime. The
$O(2)$-symmetry of the Euclidean $\gamma_{ab}$ carries over to an
$O(1, 1)$-symmetry of the Lorentzian counterpart. Therefore, the metric solution
is simply obtained by formally taking $\varphi \rightarrow i t$. The line element then reads
\begin{equation}
\dd s^2 = -\rho^2(\chi) \dd t^2 + \dd \chi^2 + R^2(\chi) \left[\dd \theta^2 + \sin^2 \theta \dd \phi^2\right]~.
\end{equation}
Using $R'^2 = -V(R) \equiv f(R)$, $\rho \propto R'$, and absorbing the constant of proportionality into the
definition of $t$, this can be written in the familiar form of Schwarzschild-type coordinates,
\begin{equation}
\label{RNdSmetric}
\dd s^2 = - f(R) \dd t^2\\ + f^{-1}(R) \dd R^2 + R^2 \left[\dd \theta^2 + \sin^2 \theta \dd \phi^2\right]~.
\end{equation}
A causal diagram of the full classical spacetime is shown in Fig.~\ref{fig:pcd}.

Strictly speaking, the analytic continuation only covers the patch $R_b < R < R_c$.
However, the metric (\ref{RNdSmetric}) can simply be extended to allow
$R > R_c$ or $R < R_b$, respectively IV and II on Fig.~\ref{fig:pcd},
in order to cover the entire physical manifold. 
In these patches, $R$ is a timelike coordinate (since $f(R)$ is negative there),
and $t$ is a spacelike coordinate, so that the metric in manifestly homogeneous 
(while it is static in the domain between the horizons). Furthermore,
using $R$ as a time coordinate one obtains a foliation
$\mathbb{R}_{\mathrm{time}} \times (\mathbb{R} \times \mathrm{S}_2)_{\mathrm{space}}$, which means
that the two patches may be considered as two separate KS universes. 

The cosmology of our own universe is described by the KS spacetime of region IV. For a
general tunneling process in the landscape of compactifications, region II will be replaced 
by a patch containing the parent vacuum, 
which may have a number of additional large dimensions, cf.\ Fig.~\ref{fig:shape-shift}.
Regions I and V then interpolate between the parent vacuum and our own universe. We assume
that this does not drastically alter the cosmology in our universe, though the early evolution
may change somewhat as the moduli describing our microscopic dimensions settle to their
new vacuum configuration.

The KS spacetime of region IV begins with a Big Bang which is a ``pancake''
singularity where only the scale factor of the non-compact direction is zero 
(since $R = R_c$ at this time). After a short period of curvature domination,
the universe starts to inflate.
In our simple model, inflation is driven by a cosmological constant
and therefore it never ends. 
This can be overcome by replacing $\Lambda$ by an inflaton field with appropriate properties.
We will assume that the initial conditions for inflaton, which are
determined by the tunneling process, allow for a period of slow-roll inflation
in the newly nucleated universe. Explicit examples for such a scenario have been
constructed in \cite{CJR}. In this case, the inflaton will eventually decay and
the universe will reheat.

We have presented a scenario of gravitational tunneling in which
our universe would have homogeneous but anisotropic 
spacelike hypersurfaces with topology 
$\mathbb{R} \times \mathrm{S}_2$. In addition to this anisotropic curvature, 
the expansion is also anisotropic.
The anisotropy of the background is reflected in observables such as
the correlation 
of multipole coefficients in the CMB. 
We now present a simplified analysis of these observable signatures.


\section{Observable signatures of the anisotropy in a Kantowski-Sachs universe}
\label{sec:perturbations}

The non-trivial topology and geometry of the background introduces
several changes with respect to Friedman-Robertson-Walker cosmologies.
First, because of the compact dimensions, the power spectrum on 
large scales is modified. In the event that 
the period of inflation was short, these scales 
would enter our horizon today.
Second, the theory of cosmological perturbations is more intricate 
because scalar, vector, and tensor perturbations are coupled via the shear
(see, e.g., \cite{GCP,PPU2,Watanabe:2010fh} for related work in Bianchi I models). 
Third, the free streaming of the photons is also anisotropic, i.e., 
the redshift factor between the last scattering and today depends on the direction
of observation (an ``anisotropic Sachs-Wolfe effect'').

These three sources of primary anisotropies 
cannot be treated independently in a self-consistent way because
they have a common origin.
We will nevertheless examine them separately 
by means of the following two simplifications.
First, we will not present the full theory of cosmological perturbations
but only examine the power spectrum of a test scalar field.
Second, we  disentangle the ``early'' from the ``late'' time effects, that is
those resulting from the modifications of the inflationary power spectrum 
and the anisotropic Sachs-Wolfe effect.
This is achieved by assuming that the expansion rate is 
isotropic after the end of inflation.

In sec.\ \ref{sec:power phi}, we solve the mode equation of a massless scalar field
in KS spacetime, motivate the choice of the state, and give an analytic expression of the 
corresponding two-point correlation function at equal time.
In sec.\ \ref{sec:power alm} , 
we project this two-point correlation function 
onto a sphere and calculate
the corresponding multipole correlators $\langle a_{lm}  a^\ast_{l'm'} \rangle$ 
for $l,l' = 2,3$ and $4$. 
In sec.\ \ref{sec:rafinements}, we discuss the additional corrections introduced by the 
remaining sources of anisotropies.
In sec. \ref{sec:anomalies} we examine whether some of the reported 
CMB anomalies can be accounted for by the model.

\subsection{Power spectrum of a massless scalar field in a Kantowski-Sachs spacetime}
\label{sec:power phi}

We begin by introducing a new set of coordinates in the inflationary KS universe
of RNdS.
It is important for the quantization and the definition of the ground state 
to work with a time coordinate 
which can be continued analytically beyond the cosmological horizons (into I and V of 
Fig.~\ref{fig:pcd}). The following choice satisfies this requirement:
\begin{equation} \label{metricmode}
\dd s^2 = H^{-2} \left[-\dd \tau^2 + V(R) \dd z^2\right] + R^2 \left[\dd \theta^2 + \sin^2\theta \dd \phi^2\right]~,
\end{equation}
where we have introduced the inflationary Hubble parameter, $H \equiv \sqrt{\Lambda / 3}$.
In this equation, $R$ should be read as a function of $\tau$. Note that $V(R)$ is positive and
plays the role of a scale factor for the non-compact direction $z$. 
As can be seen by comparing with eq.~(\ref{RNdSmetric}), 
the spacelike coordinate is defined by 
$\dd z = H\, \dd t$, and 
the (dimensionless) time coordinate by $\dd \tau = H\, \dd R/\sqrt{V(R)}$.
$R$ is thus a growing function of $\tau$.

Let us consider a minimally coupled massless real scalar field $\Phi$, 
\begin{equation}
\mathcal{S}_{\Phi} = -\frac{1}{2} \int\!\dd^4 x \sqrt{-g} g^{\mu\nu} \partial_\mu \Phi \partial_\nu \Phi~.
\end{equation}
Given a Cauchy surface of the entire spacetime
such as $\Sigma_0$ on Fig.~\ref{fig:pcd}, $\Phi$ and its conjugate momentum are 
required to satisfy the canonical commutation relations.
We note the Klein-Gordon product 
\begin{equation}
 \left( f,g \right) = i \int_\Sigma\!\!d\sigma^\mu \, 
  \sqrt{h_\Sigma} \left( f^* \partial_\mu g - g \partial_\mu f^* \right)
\end{equation}
and decompose the field into creation and annihilation 
operators $\mathrm{a}_n^\dagger, \mathrm{a}_n$ defined by
\begin{equation}
  {\mathrm{a}}_n = \left( f_n, \Phi \right)~,
\end{equation}
where $\left\{f_n\right\}$ is a complete family of positive norm solutions 
of the Klein-Gordon equation to be specified below.
Following \cite{Sasaki}, we carry out this program in the following two steps.
First we find the normalized
solutions of the field equation in the KS wedge. 
Then we analytically continue the solutions beyond the 
cosmological horizon (into regions I and V of Fig.~\ref{fig:pcd}) and 
demand that they be regular there. This way, we avoid normalizing the 
solutions of the Klein-Gordon equation in regions I and V, which would 
require the calculation of a complicated integral.

Exploiting the $\mathbb{R} \times \mathrm{S}_2$ symmetry of the spacelike hypersurfaces of KS,
we introduce the decomposition 
\begin{equation}
\label{modeexpansion}
\Phi(\tau, z, \theta, \phi) = \sum_{\ell, m} \int\!\frac{\dd k}{\sqrt{2 \pi}} \, 
 \Phi_{k\ell m}(\tau) e^{i k z} Y_{\ell m}(\theta, \phi)~.
\end{equation}
We introduce the rescaled field $\Psi = \vert g \vert^{1/4} \Phi=  R V^{1/4} \Phi$, 
in terms of which the action reads
\begin{multline}
\mathcal{S}_{\Psi} = \frac{1}{2} \sum_{\ell, m} \int\!\dd k \dd \tau 
 \biggl\{\dot{\Psi}_{k\ell m} \dot{\Psi}^\ast_{k\ell m} - \biggl[\frac{k^2}{V(R)} + \frac{\ell \left(\ell + 1\right)}{H^2 R^2}\biggr.\biggr.\\ \biggl. \biggl. - \frac{1}{H^2} \left(\frac{V'(R)}{R} - \frac{V'(R)^2}{16 V(R)} + \frac{V''(R)}{4}\right)\biggr] \Psi_{k\ell m} \Psi^\ast_{k\ell m}\biggr\}~,
\end{multline}
where we have used $\ddot{R} = V'(R) / 2 H^2$ and dropped the boundary term of an 
integration by parts.
We further decompose the solutions into
\begin{equation}
\label{ladderops}
 {\Psi}_{k\ell m}(\tau) =  {\mathrm{a}}_{k\ell m} \, u_{k\ell}(\tau) 
 + (-1)^{m}\,  {\mathrm{a}}^\dagger_{-k\ell -m}\,  u^\ast_{k\ell}(\tau)~.
\end{equation} 
where the functions $u_{k\ell}$ satisfy the equation
\begin{multline}
\label{modeeq}
\ddot{u}_{k\ell} + \biggl[\frac{k^2}{V(R)} + \frac{\ell \left(\ell + 1\right)}{H^2 R^2}\biggr.\\ \biggl. - \frac{1}{H^2} \left(\frac{V'(R)}{R} - \frac{V'(R)^2}{16 V(R)} + \frac{V''(R)}{4}\right)\biggr] u_{k\ell} = 0
\end{multline}
The mode functions defined by (\ref{modeexpansion})--(\ref{modeeq}) 
form a complete family of solutions of the Klein-Gordon equation and are 
normalized provided 
\begin{equation}
\label{Wronskian}
\mathrm{Im}\left(u_{k\ell} \dot{u}^\ast_{k\ell}\right) = 1~.
\end{equation}
The particular solution of (\ref{modeeq}) is fixed by
demanding that the analytic continuation of the 
functions $u_{k \ell}$ to the entire spacetime 
is regular on a Cauchy surface, 
say, $\Sigma_0$ of Fig.~\ref{fig:pcd} for definiteness.

In order to carry out this calculation explicitly, 
we make the approximation $V(R) \sim H^2 R^2 - 1$.
One may notice that this corresponds to work with a 
$\mathbb{R} \times \mathrm{S}_2$ foliation of 4D de~Sitter space 
(see the appendix) and one expects that it is good only if there
is a separation of scales $R_c \gg R_b, R_i$.
With this approximation, we  
have $R(\tau) = H^{-1} \cosh \tau$.
We can now eliminate $R$ from (\ref{modeeq}),
\begin{equation} \label{pseudoradial}
\ddot{u}_{k\ell}(\tau) + \left[\frac{k^2 + 1/4}{\sinh^2 \tau} + \frac{\ell \left(\ell + 1\right)}{\cosh^2 \tau} - \frac{9}{4}\right] u_{k\ell}(\tau) = 0~.
\end{equation}
This equation can be solved exactly by substituting $s \equiv \sinh^2 \tau$. 
The resulting equation is recognized as Riemann's differential equation  
whose general solution can be written in terms of the hypergeometric
function ${}_2F_1$:
\begin{multline}
\label{modesol}
u_{k\ell} = s^{1/4} s^{i k / 2} \sqrt{\left(s + 1\right)^{\ell + 1}}\\ \times \left[ {}_2F_1\!\left(\frac{\ell + i k}{2}, \frac{\ell + 3 + i k}{2}; 1 + i k; -s\right) A_{k\ell}\right.\\ \left.+ \left(-s\right)^{-i k} {}_2F_1\!\left(\frac{\ell - i k}{2}, \frac{\ell + 3 - i k}{2}; 1 - i k; -s\right) B_{k\ell}\right]
\end{multline}
for $\ell \geq 1$. Here and in the following, we place the branch cuts such that
all functions are analytic in the entire upper half of the complex plane of $s$, including
the real line but allowing for isolated singular points. 
The case $\ell = 0$ has to be treated separately, the general solution in this
case is
\begin{multline}
\label{modesol0}
u_{k0} = s^{1/4} s^{i k / 2} \left[ {}_2F_1\!\left(\frac{i k - 1}{2}, 1 + \frac{i k}{2}; 1 + i k; -s\right) A_{k0}\right.\\ \left.+ \left(-s\right)^{-i k} {}_2F_1\!\left(\frac{-i k - 1}{2}, 1 - \frac{i k}{2}; 1 - i k; -s\right) B_{k0}\right]~.
\end{multline}
The constants of integration, $A_{k\ell}$ and $B_{k\ell}$, are fixed by 
the conditions of regularity and of normalization\footnote{In a
first preprint of this paper we used a state 
corresponding to the solutions $\sim e^{-ik\eta}$, with $d\eta = d\tau/\sqrt{V[R(\tau)]}$ 
a conformal time coordinate. 
We see from the limiting form of eq.~(\ref{pseudoradial}) near the 
horizon, i.e.\ $\tau \to 0$, that they correspond to $A_{k\ell}=0$. 
Some of the integrals in (\ref{final-alm}) are, however,
infrared divergent for that spectrum. This
infrared contribution dominated the (truncated) integrals and
was responsible for a different scaling of the corrections reported there 
($\propto r$ instead of $\propto r^2$).
We wish to thank Mike Salem for helpful correspondence concerning the appropriate definition of the vacuum state.}.
To implement the former, we note that we can cover the static regions
of RNdS (I and V in Fig.~\ref{fig:pcd}) by integrating $\dd \tau = H\, \dd R/\sqrt{V(R)}$
to values $R < R_c$. The imaginary part of $\tau$ then plays the role of a radial
coordinate. In a RNdS spacetime, we would impose regularity at $R = R_b$, the ``north pole''
of the geometry. Since we are using de~Sitter space as an
approximation, we instead use 
this condition at $R = 0$, i.e.\ at the poles of Euclidean de~Sitter space 
$\tau \to\pm  i\pi/2$.
As seen in (\ref{pseudoradial}), for $\ell \geq 1$ the solutions near the poles  
asymptote to $u_{k\ell} \sim \rho^{\ell+1}$ or $u_{k\ell} \sim \rho^{-\ell}$, where
$\tau = \pm i(\pi/2 - \rho)$. The second solution is not normalizable and
is therefore excluded from the physical spectrum. 
Expanding the solution (\ref{modesol}) around $s=-1$, one sees that it remains regular if
the constants of integration fulfill the relation
\begin{equation}
B_{k\ell} = -A_{k\ell} \frac{\Gamma\left(1 + i k\right)
  \Gamma\left(\frac{\ell - i k}{2}\right) \Gamma\left(\frac{\ell + 3 -
      i k}{2}\right)}{\Gamma\left(1 - i k\right)
  \Gamma\left(\frac{\ell + i k}{2}\right) \Gamma\left(\frac{\ell + 3 +
      i k}{2}\right)}~. 
\end{equation}
For $\ell = 0$, we impose 
\begin{equation} \label{choice l=0}
\frac{\dd u_{k0}}{\dd R}\Big|_{R = 0} = 0 
\end{equation} 
such that the derivative is continuous at the poles. 
The constants of integration in (\ref{modesol0}) are therefore related by
\begin{equation}
B_{k0} = -A_{k0} \frac{\Gamma\left(1 + i k\right) \Gamma\left(\frac{-1 - i k}{2}\right) \Gamma\left(1 - \frac{i k}{2}\right)}{\Gamma\left(1 - i k\right) \Gamma\left(\frac{-1 + i k}{2}\right) \Gamma\left(1 + \frac{i k}{2}\right)}~.
\end{equation}
Using these relations, one can show that the normalization of the Wronskian, eq.~(\ref{Wronskian}), implies
\begin{equation}
\left|A_{k\ell}\right|^2 = \frac{1}{k \left(e^{2 \pi k} - 1\right)}~,
\end{equation}
including the case $\ell = 0$. This fixes the vacuum mode functions up to an irrelevant overall phase.

The two-point correlation function (at equal time) is finally given by 
\begin{multline}
\label{2ptfunction}
\left<\Phi(\tau, 0, 0, 0) \Phi(\tau, z, \theta, 0)\right> =  
\\ \sum_\ell \frac{\left(2 \ell + 1\right)}{16 \pi^2} \int\!\dd k \, 
\frac{\left|u_{k\ell}\right|^2}{R^2 \sqrt{V(R)}} P_\ell\left(\cos \theta\right) e^{-i k z}~,
\end{multline}
where we used homogeneity of space to 
shift one of the points to the origin of the coordinate system
$(z = 0, \theta = 0)$, and 
the rotational symmetry of $\mathrm{S}_2$ to set the
azimuthal coordinate $\phi$ of the second point to zero. 
We call $\left|u_{k\ell}\right|^2 / R^2 \sqrt{V(R)}$ 
the ``power spectrum''.
It becomes time-independent at late time $\tau \rightarrow \infty$,
\begin{equation} \label{power1}
\frac{\left| u_{k\ell} \right|^2}{R^2 \sqrt{V(R)}} \rightarrow \frac{H^2 \left|\Gamma\left(\frac{\ell + i k}{2}\right)\right|^2}{2 \left(k^2 + \left(\ell + 1\right)^2\right) \left|\Gamma\left(\frac{\ell + 1 + i k}{2}\right)\right|^2} \equiv \mathcal{P}_{k\ell}~,
\end{equation}
for $\ell > 0$, and
\begin{equation} \label{power2}
\frac{\left| u_{k0} \right|^2}{R^2 \sqrt{V(R)}} \rightarrow \frac{H^2 \tanh\frac{\pi k}{2}}{k + k^3} \equiv \mathcal{P}_{k0}~,
\end{equation}
for $\ell = 0$.
These results are most conveniently obtained by taking the limit $s \rightarrow \infty$ of eqs.~(\ref{modesol})
and (\ref{modesol0}), and noting that $R^2 \sqrt{V(R)} \rightarrow H^{-2} s^{3/2}$.


\subsection{Multipole coefficients}
\label{sec:power alm}

The CMB anisotropies are described by 
the two-point correlation function 
on the two-dimensional intersection of the past light-cone of the observer
with a constant time hypersurface.
Anisotropic expansion during the matter dominated era causes 
an angular dependent redshift but for simplicity 
we shall neglect this additional corrections to the standard angular power spectrum
in this section.
Namely, we evaluate the two-point correlation function 
given in the previous section on a 
surface at fixed comoving distance 
and calculate the correlations between the multipole coefficients 
$\left< a_{lm} a^\ast_{l'm'}\right>$. 
Our approximation amounts to assume that the post-inflationary phase of expansion is isotropic,
\begin{equation}
\label{latemetric}
\dd s^2 \sim - \dd t^2 + R^2(t) \left[\dd z^2 + \dd \theta^2 + \sin^2 \theta \dd \phi^2\right]~.
\end{equation}

We can choose the coordinate system so that the observer is at the origin 
$(z = 0, \, \theta = 0)$. 
The direction of observation $\mathbf{n}$ defines a point 
$(z = r \, \mathbf{z}\cdot\mathbf{n}, \, \theta = r\, \sqrt{1 - (\mathbf{z}\cdot\mathbf{n})^2})$
on the last scattering surface, 
where $\mathbf{z}$ is the unit vector pointing along the $z$-direction, and $r$ is the comoving 
radius of the last scattering surface,
\begin{equation} \label{important}
  r = \int_{t_{\rm rec}}^{t_0}\!\!\frac{dt}{R(t)} \simeq \frac{3.5}{\dot{R}(t_0)} = 
  3.5\, \sqrt{\Omega_{\rm curv}}~,
\end{equation} 
see the appendix for the definitions.
We also record the relation between the ratio $r$ 
and the number of $e$-folds of inflation. 
Since the shear is small, the 
radius of the last scattering surface is in the first approximation equal to 
its value in isotropic cosmologies $l_{\rm LSS} \simeq 0.5 H_0^{-1}$.
Let us define the origin of the number of e-folds as the instant where 
the radius of the compact directions $R$ is of the order of the Hubble radius today
$H_0^{-1} \simeq R$.
If inflation lasts $N_{\rm extra} > 0$ 
additional $e$-folds, this 
radius is $e^{N_{\rm extra}}$ larger. Hence 
\begin{equation} \label{r}
  r = \frac{l_{\rm LSS}}{R} \simeq 0.5 \, e^{-N_{\rm extra}}~.
\end{equation}
A long period of inflation therefore corresponds to $r \ll 1$.

The correlations between the multipole coefficients are given by
\begin{equation} \label{correlations a_lm}
\left< a_{lm} a^\ast_{l'm'} \right> = \int \!\dd \Omega \dd \Omega' Y_{lm}(\mathbf{n}) Y^\ast_{l'm'}(\mathbf{n}') \left< \Phi(\mathbf{n}, r) \Phi(\mathbf{n'}, r) \right>~.
\end{equation}
Before we start to calculate, 
we note that some of them vanish by invariance under point-reflection,  
which is a symmetry of the KS model.
The two-point function
in the above expression is
indeed even under parity and 
the correlations between multipole coefficients of opposite parity,
i.e.\ such that $l+l'$ is odd, vanish identically.
Parity is also a symmetry of the Bianchi III models considered in
\cite{BPS} and \cite{GHR}, 
which explains why they also find no correlations for the multipole coefficients
where $l+l'$ is odd.

We now introduce intrinsic spherical coordinates $(\vartheta, \varphi)$ 
on the last scattering surface.
For simplicity, we choose the polar axis aligned with the $z$-direction
such that $\mathbf{z}\cdot\mathbf{n} = \cos\vartheta$.
As we have already seen in the previous section, 
homogeneity of space guarantees that the two-point function
in the above equation only depends on two parameters: the separation 
along the flat direction $z$, given by 
$r (\cos \vartheta - \cos \vartheta')$, 
and the angle $\theta$ subtended by 
an arc on the compact dimensions, that is
$\cos \theta = \cos (r \sin \vartheta) \cos (r \sin \vartheta') + 
\sin (r \sin \vartheta) \sin (r \sin \vartheta') \cos (\varphi - \varphi')$, 
as some simple geometric
considerations show. In order to simplify our expressions, we will use the identity
\begin{equation}
P_\ell\left(\cos \theta\right) = \frac{4 \pi}{2 \ell + 1} \sum_{n = -\ell}^{\ell} Y^\ast_{\ell n}(r \sin \vartheta, \varphi) Y_{\ell n}(r \sin \vartheta', \varphi')~.
\end{equation}
Using eq.~(\ref{2ptfunction}),
the multipole correlations in terms of these intrinsic coordinates become
\begin{multline}
\left< a_{lm} a^\ast_{l'm'} \right> = \int \!\dd \Omega \dd \Omega' \sum_\ell \int\!\dd k Y_{lm}(\vartheta, \varphi) Y^\ast_{l'm'}(\vartheta', \varphi') \\ \times \frac{\mathcal{P}_{k\ell}}{4 \pi} \sum_{n = -\ell}^{\ell} Y^\ast_{\ell n}(r \sin \vartheta, \varphi) Y_{\ell n}(r \sin \vartheta', \varphi')\\ \times e^{-i k r \left(\cos \vartheta - \cos \vartheta'\right)}~.
\end{multline}

Since KS spacetime is axially symmetric around the $z$-axis, and since 
our choice of intrinsic coordinates respects that symmetry, 
the coefficients $a_{lm}$ with different $m$ are uncorrelated. 
One can show this explicitly by integrating
out the angles $\varphi$ and $\varphi'$,
\begin{widetext}
\begin{multline} \label{final-alm}
\left< a_{lm} a^\ast_{l'm'} \right> = \delta_{mm'} \sqrt{2 l + 1} \sqrt{2 l' + 1} \sqrt{\frac{\left(l - m\right)!}{\left(l + m\right)!} \frac{\left(l' - m\right)!}{\left(l' + m\right)!}} \int_0^\pi\!\sin\vartheta \dd \vartheta \int_0^\pi\!\sin\vartheta' \dd \vartheta' \int\!\dd k P_{lm}(\cos\vartheta) P_{l'm}(\cos\vartheta')\\ \times \sum_\ell \mathcal{P}_{k\ell} \frac{2 \ell + 1}{16 \pi} \frac{\left(\ell - m\right)!}{\left(\ell + m\right)!} P_{\ell m}\Bigl(\cos (r \sin \vartheta)\Bigr) P_{\ell m}\Bigl(\cos (r \sin \vartheta')\Bigr) e^{-i k r \left(\cos \vartheta - \cos \vartheta'\right)}~.
\end{multline}
\end{widetext}

We calculate the remaining coefficients 
$\left< a_{lm} a^\ast_{l'm'} \right>$ numerically.
We compare these results with the statistically
isotropic case of a flat Friedman-Robertson-Walker universe with a 
Harrison-Zel'dovich spectrum 
\begin{equation} \label{FRWresult}
  \left< a_{lm} a^\ast_{l'm'} \right>_\mathrm{iso} 
  = \frac{H^2}{2 \pi} \, \frac{1}{l \left(l + 1\right)}\, \delta_{ll'} \delta_{mm'} ~.
\end{equation}
This expression is obtained by computing the vacuum expectation value in 
the Bunch-Davies vacuum using the flat slicing of de~Sitter space.

\begin{figure}[t]
\includegraphics[width=85mm]{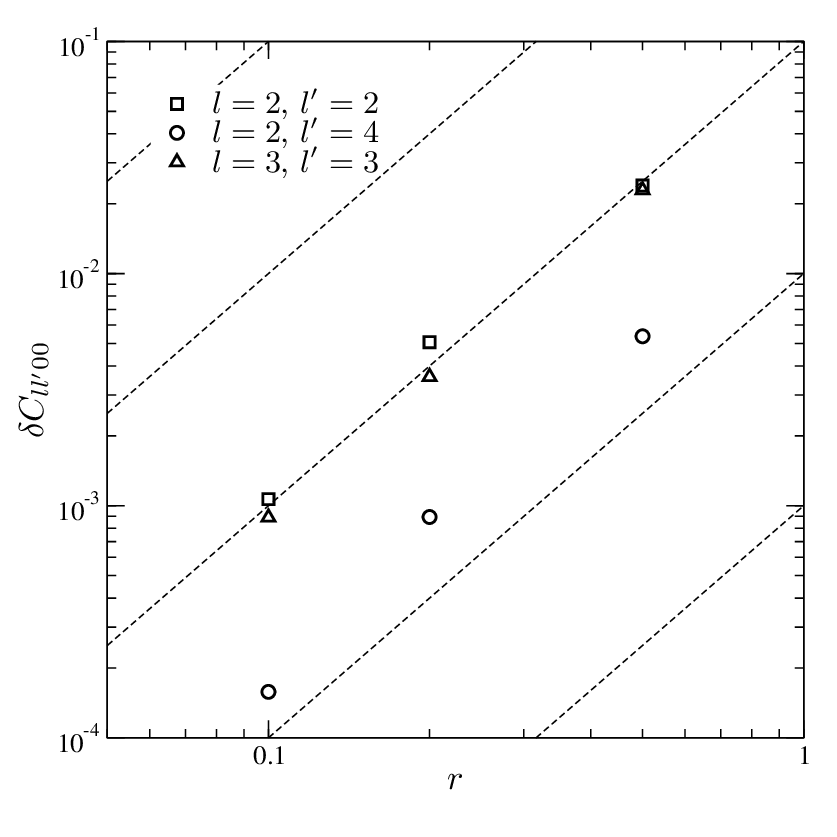}
\caption{\label{fig:scaling} \small Amplitude of the anisotropic correction to the
multipole correlators for different values of $r$ (three examples). 
We found a scaling consistent with 
$\delta C_{ll'mm'} \propto r^2$, as indicated by the dashed lines.}
\end{figure}

We present our results in the following form
\begin{multline} \label{almvariance}
\left< a_{lm} a^\ast_{l'm'} \right> = \frac{H^2}{2\pi} 
  \delta_{mm'} \left(\delta_{ll'} + \delta C_{ll'mm}\right) \\
\times \max\left\{\frac{1}{l \left(l + 1\right)}, \frac{1}{l' \left(l' + 1\right)}\right\}~,
\end{multline}
and give the coefficients $\delta C_{ll'mm}$ in table 
\ref{tab:numerics} for the case $r=0.5$,
which corresponds to $\Omega_{\rm curv} \simeq 0.02$, see
eq.\ (\ref{important}). The scaling of the coefficients $\delta C_{ll'mm}$
with $r$ is illustrated in Fig.~\ref{fig:scaling}.
Our findings are the following.
First, the amplitude of the corrections is suppressed by $r^2 \propto \Omega_{\rm curv}$.
In the limit of a long period of inflation, $r \rightarrow 0$, 
i.e., when the observer only has access to scales much smaller 
than the radius of $\mathrm{S}_2$, the anisotropies become unobservable.
Second, as already pointed out the correlations vanish by parity for $l+l'$ odd. 
This result does not depend on our choice of coordinate system.
Third, the amplitude of the corrections is largest for $l=l'$, and generally decreases
with growing $\left|l - l'\right|$. It is also noteworthy that the amplitude, at fixed $\left|l - l'\right|$,
is almost independent of $l$, which means that the correlations extend
uniformly up to arbitrarily high multipoles.
Fourth, for $l=l'$, the $m$-dependence of the corrections obeys
\begin{equation} \label{parabola}
\delta C_{llmm} = \delta C_{ll00} \left(1 - \frac{3 m^2}{l \left(l + 1\right)}\right)
\end{equation}
to a good approximation. 
As a result, the corrections cancel in the averaged multipole 
\begin{equation} \label{C_l}
  C_l \equiv \sum_m \frac{\left<a_{lm} a^\ast_{lm}\right>}{2 l + 1} 
= \frac{H^2}{2\pi } \frac{1}{l(l+1)}~.
\end{equation}
Fifth, from eq. (\ref{important}) we deduce
\begin{equation} \label{amplitude}
  \delta C_{ll00} \simeq \Omega_{\rm curv}\, .
\end{equation}
A mild dependence on $l$ of the prefactor is possible.

We emphasize two remarkable properties of these spectra.
The first one is the scale invariance of the power spectrum, and 
in particular the absence of an infrared cutoff  
which one could have naively expected from the asymptotic form of the power spectrum 
(\ref{power1}) and (\ref{power2}) at $k\to 0$ or by analogy with
closed FRW models. 
The second is the fact that, as we already pointed out, at a given value of $l$, 
the anisotropy manifests itself by a redistribution of the power amongst the 
various $a_{lm}$ according to the ``sum rule''
\begin{equation} \label{sum rule}
  \sum_m \delta C_{llmm} = 0\, .
\end{equation}
In sec.\ \ref{sec:anomalies}, we show how eqs.\ (\ref{parabola}) and 
(\ref{sum rule}) translate into observables (the so-called bipolar coefficients).
Equations (\ref{parabola})-(\ref{sum rule}) and table \ref{tab:numerics} 
are the central results of this section.\footnote{We also note 
that because of the sum rule (\ref{sum rule}),
the corrections to the (cosmic) variance of the estimator 
$\hat{C}_l \equiv \sum_m \left|a^{\rm obs}_{lm}\right|^2 / (2 l + 1)$
are only quadratic in $\Omega_{\rm curv}$. 
Indeed with use of eq.~(\ref{parabola}) one obtains
\begin{equation}
\mathrm{Var}\left[\hat{C}_l\right] = \frac{2 C_l^2}{2 l + 1} \left[1 + \frac{1}{5} \left(4 - \frac{3}{l \left(l + 1\right)}\right) \delta C^2_{ll00}\right]~, \nonumber 
\end{equation}
where it is still assumed that the field is Gaussian.}

\begin{table}[tbh]
\begin{tabular}{|@{\hspace{3.5pt}} c @{\hspace{3.5pt}} | @{\hspace{3.5pt}} c @{\hspace{7.5pt}} c @{\hspace{7.5pt}} c @{\hspace{7.5pt}} c @{\hspace{3.5pt}}|}
\hline
$\delta C_{ll'mm}$ & $m = 0$ & $m = 1$ & $m = 2$ & $m = 3$\\
\hline
$l = 2$, $l' = 2$ & {\small $+0.02(40)$} & {\small $+0.01(16)$} & {\small $-0.02(48)$} & ---\\
$l = 2$, $l' = 3$ & {\small $0$} & {\small $0$} & {\small $0$} & ---\\
$l = 2$, $l' = 4$ & {\small $-0.005(4)$} & {\small $-0.005(0)$} & {\small $-0.003(3)$} & ---\\
$l = 3$, $l' = 3$ & {\small $+0.02(29)$} & {\small $+0.01(62)$} & {\small $-0.00(04)$} & {\small $-0.02(69)$} \\
$l = 3$, $l' = 4$ & {\small $0$} & {\small $0$} & {\small $0$} & {\small $0$} \\
\hline
\end{tabular}
\caption{\label{tab:numerics} \small Relative deviations 
from a flat Friedman-Robertson-Walker cosmology of the multipole correlations.
The numerical results are for $r = 0.5$. Solid convergence was achieved for the leading digit only.
We found that for the nonzero entries
$\delta C_{ll'mm'}$ is roughly proportional to $r^2$, cf.\ Fig.~\ref{fig:scaling}.}
\end{table}

In \cite{BPS}, 
a table similar to our table \ref{tab:numerics} can be found for a Bianchi III model. 
The deviations due to anisotropy are controlled in that case by 
the comoving radius of the last scattering surface,
$\rho_\star$ in their notation, and scale like
$\rho_\star^2 \propto \Omega_{\rm curv}$. 
Their table and ours are remarkably
similar\footnote{It is worth pointing out that their calculations differ significantly from ours.
While we calculated the integrals (\ref{final-alm}) numerically, the authors of \cite{BPS} made
a series of approximations to simplify the corresponding expression 
(their eq.~($4.33$)) before resorting to numerics. 
The agreement between our 
results and theirs, where they overlap, gives a good indication that these 
approximations, rather hard to justify rigorously, are correct.},
the only difference appears to be the overall sign of the corrections. 
It is tempting to 
conjecture that this is related to the opposite sign of the
curvature of Bianchi III models as compared to KS models. 
Note as well that eq.~(\ref{parabola}) then should hold for both cases,
as seems to be implied also by Fig.\ 3 of \cite{BPS}. The resulting
cancellation of the corrections when taking the sum over $m$ was,
however, not reported in \cite{BPS}.

\subsection{Corrections to the Sachs-Wolfe effect}
\label{sec:rafinements}

Our analysis of anisotropic signatures was simplified
in two respects. 
First, we only consider a test field. Without a detailed analysis,
it is hard to determine whether the theory of cosmological perturbations 
in KS spacetime would give primordial spectra 
qualitatively different from the ones of perturbed FRW spacetimes, besides 
the ones already found for test scalar fields. 
We leave this question for future work.
Our second simplification is that we neglected the anisotropic redshift
after last scattering. In a KS spacetime, this effect is 
controlled by the ratio of the shear, $\delta$, to the isotropic expansion rate $H$
(see the appendix for the definitions).
This ratio decreases during inflation but grows 
again during radiation and matter domination, 
so that it becomes significant only at small redshifts $z = \mathcal{O}(1)$. 
These late time effects are thus determined by the value of the ratio
\begin{equation}
  \sigma = \frac{\delta_0}{H_0}\,\,.
\end{equation}

To get a handle on the additional corrections brought by the 
anisotropic redshift of photons since last scattering,  
note that this effect has two origins in our model. One is caused by the anisotropic 
expansion and the second by the curvature of the homogeneous surfaces.
To see the effect of the anisotropic expansion alone, 
consider a Bianchi I model  
$ds^2_{\rm B \, I} = -dt^2 + a^2(t)dx^2 + b^2(t)(dy^2 + dz^2)$. 
The null geodesic equation is readily solved and gives the temperature distribution
in the direction $(\vartheta,\varphi)$ 
\begin{equation} \label{ellipse}
  T_{\rm B \, I} (\vartheta,\varphi,t_O) = T_{\rm em} \, 
\left( \frac{x^2}{a_{\rm em}^2} + \frac{y^2 + z^2}{b_{\rm em}^2} \right)^{-1/2}
\end{equation}
where $T_{\rm em}$ is the temperature of the Planck spectrum at the time of emission, 
and the cartesian coordinates $(x,y,z)$ are related to the intrinsic coordinates
on the unit sphere by $x=\sin \vartheta \cos \varphi$, 
$y=\sin \vartheta \sin \varphi$, and $z=\cos \vartheta$.
The quadrupole character of this distribution is obvious
(recall that when written in cartesian coordinates, the 
spherical harmonics of weight $l$ are homogeneous polynomials of degree $l$).
We conclude that the anisotropic Sachs-Wolfe effect takes the form
\begin{equation} \label{quadmodulation}
  \left( \frac{\delta T}{T}\right)_{\rm B \, I} 
  \sim \frac{\Psi}{3} \left( 1 + 
  \sum_{m=-2}^{m=2} \mathcal{O}(\sigma) Y_{2m}(\vartheta,\varphi) \right)
  + \ldots
\end{equation} 
where $\Psi$ is the gravitational potential in the Newtonian gauge and 
the ellipsis stands for the corrections from the velocity potential of matter and
the integrated Sachs-Wolfe term, both of which receive similar quadrupolar 
{\it modulations}.
Inserted into the expression (\ref{correlations a_lm}) of the correlation functions, 
this quadrupole generates additional correlations between the 
$\left< a_{lm} a^\ast_{l'm'} \right>$ with $l' = l, l\pm 2$.

The non Euclidean character of the surface of homogeneity 
in Kantowski-Sachs spacetimes 
is responsible
for additional corrections to the geodesic equation. As a result, the 
temperature distribution cannot be written in closed form as in 
the simple expression (\ref{ellipse}).
Nevertheless, inspection of the geodesic equations is sufficient to show 
that these corrections scale with $\sigma$ as one could expect. 
Indeed, the null component of the momentum is given by
\begin{equation}
  \frac{dp_0}{d\lambda} = \delta \left\{ - 2 e^{4\Delta} \bar p_z^2 + 
  e^{-2\Delta} \left( \bar p_\theta^2 + \frac{\bar p_\phi^2}{\sin^2 \theta} 
 \right)\right\}
\end{equation}
where $\Delta = \int^t\!\!\delta$ and $\bar p_{z,\theta,\varphi}$ are integration constants.

A detailed calculation of the bolometric flux, which includes 
all of these corrections, was done in \cite{GHR}
for Bianchi III spacetimes. It can easily be adapted to the Kantowski-Sachs spacetime
and confirms the qualitative results we have just derived.
Moreover, it is noteworthy that the diagonal corrections $\delta C_{llmm}$ 
retain the same parametric form (\ref{parabola})
and therefore verify the sum rule (\ref{sum rule}), 
as can be seen from eq.~(47) of \cite{GHR}, with
the use of their equations (B1) and (B3).
Since we expect that
both early and late modifications of the power spectra 
add at leading order in $\Omega_{\rm curv}$, 
the total correction should therefore also verify 
eqs.~(\ref{parabola}) and (\ref{sum rule}).


\subsection{CMB anomalies}
\label{sec:anomalies}

We now ask whether the correlations of the multipole coefficients 
we have computed in the previous sections can account for some of the anomalies found 
in the CMB (see \cite{WMAP7anomalies, Copi:2010na} for recent reviews and references).
We can, in fact, discuss both Kantowski-Sachs (KS) and Bianchi III (BIII) models  
because 
they have important features in common and therefore make very similar predictions.
Moreover, although we have not calculated explicitly the corrections from 
the recent history of the universe, 
we saw that the parametric form eq.~(\ref{parabola}) holds for both, and 
therefore for the total correction of the CMB spectra.

Here we make no statement regarding either the cold spots, which would require an 
analysis far beyond the scope of this paper, or the alignment of the 
dipole and quadrupole with the ecliptic, which is unlikely to have a 
cosmological explanation.

Also, from the scale invariance of the 
primordial power spectrum, and in particular the 
absence of a cutoff (see eq.~(\ref{C_l}))
we conclude that the KS and BIII models
seem unable to explain the lack of power at large scales (of the quadrupole and 
of the correlation function) and  
the disparity between odd and even values of $l$.
However, the quadrupole receives a non-stochastic contribution
in anisotropic models which, in a complete calculation, may have a number
of different sources (see sec.\ \ref{sec:bound} for further details.)

The KS and BIII models can, however, in principle account for 
the alignment of the quadrupole and octopole and 
the quadrupolar power asymmetry.
A statistical analysis reveals that the effect of anisotropy on the 
alignment is not significant. On the other hand,
we find good indications that the quadrupolar asymmetry can be explained by 
the KS model, but not by BIII.
However, as we now explain the amount of anisotropic curvature needed 
to produce a signal with the observed
amplitude, $\Omega_{\rm curv} \simeq 10^{-2}$, is probably inconsistent with the non-detection
of a CMB quadrupole at the same level.

\subsubsection{The quadrupole moment and the bound on curvature}
\label{sec:bound}

In anisotropic models, the late-time expansion couples 
multipoles $l$ and $l+2$ \cite{GHR}. 
The quadrupole thus receives a {\it non-stochastic} contribution 
from the monopole proportional to $T_0 \Omega_{\rm curv}$. 
Given a model with a scale invariant power spectrum of amplitude
$\delta T / T_0 \sim 10^{-5}$, and in the absence of other sources of quadrupolar 
anisotropy, the value of anisotropic curvature is thus expected to be \cite{boundOmega}
\begin{equation} \label{bound}
  \Omega_{\rm curv} \lesssim 10^{-4}~.
\end{equation}
For the purpose of model building we note that, through the relationships 
(\ref{important}) and (\ref{r}), 
it implies a lower bound on the duration of inflation, $N_{\rm extra} \gtrsim 2.7$.
Moreover, we will see that this bound, if correct, is too stringent
for the model to account for the quadrupolar asymmetry in the CMB.

It is, however, important to stress that this bound is a rough estimate
obtained under the assumption of adiabatic perturbations \cite{boundOmega}.
A detailed calculation might show this estimate to be too naive, since
compensations might occur between the Doppler,
proper Sachs-Wolfe and integrated Sachs-Wolfe terms
in a KS and BIII universe,
as is the case in, e.g., open inflation scenarios
\cite{GZopen1,GZopen2}.
Moreover, a mixture of adiabatic and isocurvature perturbations presumably also
modifies the bound (\ref{bound}).
Thus one must bear in mind that eq.~(\ref{bound})
depends on the details of the model and might be too stringent.

\subsubsection{Alignment of the quadrupole and octopole}

We study the problem of the alignment with the help 
of the so-called multipole vectors \cite{multipolevectors}.  
The two quadrupole vectors define a plane whose normal $\mathbf{v}$ is obtained
by taking the cross-product of the two vectors. 
Similarly, all possible pairings of the three octopole vectors define
three different planes, with normals $\mathbf{w}^{\left(1\right)}$, $\mathbf{w}^{\left(2\right)}$,
$\mathbf{w}^{\left(3\right)}$. 
One says that the octopole is planar when the three $\mathbf{w}^{\left(i\right)}$ are 
almost collinear, 
and one says that the quadrupole and octopole are aligned when 
they are also almost collinear with $\mathbf{v}$.

The degree of alignment is measured by the statistics
\begin{equation}
S \equiv \frac{1}{3} \sum_i |\mathbf{v} \cdot \mathbf{w}^{\left(i\right)}|
\end{equation}
A large value of $S$ indicates an alignment.
The observed value 
(taken from the Doppler-corrected ILC maps, see \cite{Sarkar:2010yj}) 
is $S_{\rm ILC7} = 0.736$. \footnote{In order to be able to compare 
our results with others, we followed the 
general practice where the vectors 
$\mathbf{v}$ and $\mathbf{w}^{\left(i\right)}$ are {\it not} 
normalized. The use of normalized vectors instead 
affects the significance level by an order of magnitude.
It was argued that the effective weighting introduced by non normalized vectors 
accounts for ``how well'' the plane is defined by the two vectors of the cross product, 
a badly defined plane being one where the angle between these 
vectors is small 
(since the norm of the vector is proportional to the sine of this angle) \cite{Copi:2010na}. 
This position is untenable, because a plane 
is geometrically defined by two non-collinear vectors. 
Hopefully, the bias introduced by the weighting is less important for the relative 
statistical significance, i.e.\ when comparing two models, than 
it is for the absolute statistical significance, i.e.\ for a given model.}

For a Gaussian and statistically isotropic distribution of $a_{lm}$
we found that only about $1\%$ 
of the realizations have a value of $S$ larger than $S_{\rm ILC7}$. 
According to \cite{WMAP7anomalies}, this ``remarkable degree of alignment''
lacks a compelling theoretical explanation. 
Unfortunately, we will now show that the class of KS and BIII models do not 
provide one.

For each model and each set of parameters, we generated $10^{6}$ independent 
realizations of a quadrupole and octopole.
To this end, the coefficients $a_{lm}$ are treated as independent 
(remember that the correlations between quadrupole
and octopole vanish) Gaussian random variables with zero 
mean\footnote{Since the late-time expansion in anisotropic models couples 
multipoles $l$ and $l+2$ \cite{GHR}, the quadrupole receives a
non-stochastic contribution 
from the monopole proportional to $T_0 \Omega_{\rm curv}$. 
We limit however our analysis to whether the correlations of the 
primordial spectrum can themselves favor an alignment.}
and variance 
$\left< a_{lm} a^\ast_{lm} \right>$
given in eq.~(\ref{almvariance}). For each realization, we calculate 
the corresponding multipole vectors from the coefficients
$a_{lm}$ using the algorithm of \cite{Weeks:2004cz}, and compute from 
them the value of $S$. 
We set $\delta C_{3300} \simeq \delta C_{2200}$, as indicated from our table 
\ref{tab:numerics}, and with the use
of (\ref{parabola}) we then study how the distribution of $S$ 
varies as a function of the single parameter 
$\delta C_{2200} \simeq\Omega_{\rm curv}$.

For both KS and BIII spacetimes
(that is for either sign of $\delta C_{2200}$), the mean of the
distribution shifts towards larger values, indicating a tendency of the 
quadrupole and octopole to align.
However, the distribution of $S$ turns out to be fairly robust.
In particular, the mean value and variance of $S$ change very little.
For instance, in KS 
the mean value of $S$ increases by only
$\sim 1\%$ for $\delta C_{2200} \simeq 0.4$.
The effect is about twice as large for BIII.
The skewness depends more strongly on $\delta C_{2200}$, 
and in particular on its sign.
It increases for $\delta C_{2200} < 0$ (BIII) and decreases for
$\delta C_{2200} > 0$ (Kantowski-Sachs). For $\delta C_{2200} \simeq \pm 0.2$
the skewness changes by $\sim 20 \%$ in both cases.
As a result of the increase of both the mean and skewness,  
the number of realizations which exceed the threshold $S_{\rm ILC7}$ in BIII is increased by
$\sim (5 \%, 18 \%, 62 \%)$ for $\delta C_{2200} \simeq (-0.1, -0.2, -0.4)$, while it hardly changes
for KS.
The number of realizations nevertheless remains of the same order of magnitude as 
in the isotropic case.

In conclusion, we find that the modified multipole correlations in KS and BIII 
are unable to account for the alignment 
of the quadrupole and octopole at a statistically significant level.

\subsubsection{Quadrupolar power asymmetry}

A practical way to check whether the two-point correlation function
${\cal C}(\mathbf{n},\, \mathbf{n}')$ is invariant under rotation is to consider its 
representation in the basis of the total angular momentum 
operator of eigenvalues $L$ and $M$ 
(sometimes referred to as bipolar spherical harmonics \cite{Hajian:2003qq}). 
Using the `bra-ket' notation, we have
\begin{equation} \label{totangmoment}
  \langle \mathbf{n}, \mathbf{n}' \,\vert \,  {\cal C} \rangle = 
   \sum_{l,l',L,M} N^L_{ll'} A^{LM}_{ll'} \, 
   \langle \mathbf{n}, \mathbf{n}'\, \vert l l'; L M  \rangle~,
\end{equation} 
where we follow the normalization 
convention chosen by the WMAP team \cite{WMAP7anomalies} in order to facilitate comparison:
\begin{equation}
N^L_{ll'} \equiv \sqrt{\frac{\left(2 l + 1\right) \left(2 l' + 1\right)}{2 L + 1}} \,
 \langle l, 0, l', 0 \vert l l';L 0\rangle~.
\end{equation}
The $A^{LM}_{ll'}$ are the sum over $m$ and $m'$ of the 
$\left< a_{lm} a^\ast_{l'm'} \right>$ weighted by 
the appropriate Clebsch-Gordan coefficients.
Since the eigenstates $\vert l l'; L M \rangle$ generate a $(2 L+1)$-dimensional representation
of the rotation group, the condition that ${\cal C}$ is statistically isotropic reads 
\begin{multline}
  {\cal C}({\cal R}\mathbf{n},\, {\cal R}\mathbf{n}') = {\cal C}(\mathbf{n},\, \mathbf{n}') \quad \forall \, \mathcal{R} \in SO(3) \\
\quad \Leftrightarrow \quad A^{LM}_{ll'} = 0 \quad \forall \, L>0 ~, 
\end{multline}
where ${\cal R}$ is an arbitrary rotation.
If some of the $A^{LM}_{ll'}$ are non-zero for $L=1$, the CMB temperature presents 
a dipole modulation. If some coefficients are non-zero for $L=2$, the temperature
is modulated by a quadrupole as, for example, we found for the late time effects
eq.~(\ref{quadmodulation}) 
(see the appendix of \cite{WMAP7anomalies} for other illustrations).

Note that the $A^{LM}_{ll'}$ are in general not independent of the choice of a coordinate system.
The WMAP team uses a preferred coordinate system in which all coefficients with $M \neq 0$ are
consistent with zero. In KS spacetime this coincides with the coordinate system we use in our calculations.

With the use of eq.~(\ref{parabola}) we
find that $A^{00}_{ll} = C_l$ and
\begin{equation} \label{modulation}
A^{20}_{ll} = \frac{C_l}{\sqrt{5}} \left(4 - \frac{3}{l \left(l + 1\right)}\right) \delta C_{ll00}~.
\end{equation}
There is no dipole modulation because of the parity symmetry of the background 
which entails vanishing correlations for $l' = l \pm 1$.
The important remark concerning (\ref{modulation}) is that these coefficients are 
proportional to $C_l$. Therefore they should exhibit acoustic oscillations as well.

The WMAP team found  
that $A^{20}_{ll} \approx -2 A^{20}_{l-2,l}$.
As illustrated in the appendix of \cite{WMAP7anomalies}, this 
relation occurs when the anisotropy does not 
induce any correction to the average power spectrum but only redistributes
the power amongst the $a_{lm}$.
So we expect to find it in our model.
Having calculated the correlations only up to $l=4$, we can only check the relation for 
that value. 
From our table \ref{tab:numerics} 
we obtain $A^{20}_{24} \simeq -0.001\times H^2 / (2 \pi)$. 
With $\delta C_{4400} \simeq \delta C_{3300}$ we find indeed 
that $A^{20}_{44} \approx -2 A^{20}_{24}$.

In conclusion, the anisotropy in the KS and BIII models shows 
all of the characteristic features found in the WMAP data. 
In particular, the appearance 
of the acoustic peak in the $A_{ll}^{20}$ naturally follows from such models.
The predictions of the models however differ in the {\it sign} 
of the bipolar coefficients $A^{20}_{ll}$
which is determined by the sign of the curvature.
We find that KS predicts the correct sign.

Unfortunately, if the bound on anisotropic curvature (\ref{bound}) is correct, the 
predicted amplitude is two orders of magnitude too small.
Indeed, from Fig.\ $16$ of \cite{WMAP7anomalies}, we get
$A^{20}_{ll}/C_l \simeq 4 \cdot 10^{-2}$, which according to expression 
(\ref{modulation}) corresponds to $\delta C_{ll00}$, and therefore 
$\Omega_{\rm curv}$, of the order of a few percents.
However, as we recalled in section \ref{sec:bound} 
the anisotropic curvature has another effect: it adds a
non-stochastic component to the quadrupole of the order of $T_0 \Omega_{\rm curv}$ 
\cite{boundOmega,GHR}, which leads to the 
bound $\Omega_{\rm curv} \lesssim 10^{-4}$.
With a value of the anisotropic curvature in that range, the 
quadrupolar anisotropy cannot be accounted for by the model.

\section{Summary and conclusions}
\label{sec:conclusions}

In the landscape of string theory, vacua with differing numbers of
macroscopic dimensions and topologies  
are connected by tunneling events. In this paradigm, our universe has
been produced as the last step in a potentially long chain of events,
by tunneling from a parent vacuum with a greater, equal, or smaller
number of macroscopic dimensions. Like \cite{GHR, BPS} which appeared
when our work was near completion, we consider the latter
case. Our approach differs in the specific model under examination and
in its embedding into a cosmological scenario, which we dubbed the
``shapeshifting universe''.
Specifically, we take the example of a spacetime where two of
our current macroscopic dimensions are compactified on a two-sphere by
a magnetic flux.

The existence of a long-lived parent vacuum with two macroscopic
dimensions, while all others are stably compactified, is not a necessary
requirement for this scenario.
All we need is a precursor with two of our four large dimensions
compactified while other, currently compact, directions may
well have been macroscopic. This allows a direct connection of
anisotropic cosmologies produced by decompactification with the vast
possibilities of transdimensional vacuum transitions described in
\cite{CJR, BSV} and, 
furthermore, admits histories with greatly varying
effective values of $\Lambda$. In one particular possible scenario,
our parent vacuum was a $\mathrm{dS}_D \times \mathrm{S}_2$ spacetime
with higher $\Lambda_{\rm eff}$ and different compact directions than
our universe, connected to ours by a combined spontaneous
compactification of
$D-2$ dimensions which are now small and
decompactification of the $\mathrm{S}_2$. We have not explicitly
constructed the instanton describing this shapeshifting event but we
know of no reasons why it shouldn't exist. It is also plausible that
models can be constructed where inflation is triggered by the
transition, as described in \cite{CJR}. 

This line of investigation raises a number of new and interesting
questions. In the context of the cosmological measure problem, the
contribution of regions with different numbers of macroscopic
dimensions and transitions in between is still largely unexplored (see
\cite{CJR,BSV,BPS} for some ideas). We hope to return to this question
in future work. From the phenomenological point of
view, on the other hand, decompactification offers a concrete
framework for parameter studies of anisotropic cosmological
models. Our ability to detect statistical anisotropy hinges on the
assumption of a sufficiently short period of inflation, which again
may find some justification in the string theory landscape. Given the
observational constraints, a small but nonzero window for anisotropic
curvature remains (see, e.g., \cite{GHR}). Precisely how strong the
constraints already are, 
and how much can be gained by combining future large-scale galaxy
surveys and CMB data, can only be answered by a more detailed
investigation of the cosmological signatures of models like the one
proposed here.
It is also interesting to note that having embedded the anisotropic universe model
in a more complete scenario of quantum tunneling, where the KS spacetime corresponds 
to only a wedge of a large spacetime, solves the ambiguity in the ground state 
(compare with the discussions of \cite{GCP,PPU2}).

Our preliminary analysis of the deviations from statistical isotropy
of the CMB
shows the following trends.
Some of the off-diagonal correlations $\langle a_{lm} a_{l'm'}^* \rangle$, which vanish
in the isotropic case, are now non-zero. 
Because of the invariance under point reflection of the background, 
the correlations with $l$ and $l'$ of different parities vanish.
The nonzero correlations
receive contributions from both the inflationary phase 
and the recent expansion of the universe. Both scale as $\Omega_{\rm curv}$ in the observations.
The central results are the table \ref{tab:numerics} and 
the equations (\ref{parabola}) and (\ref{sum rule}).
We showed that as a direct consequence of this, and in contrast to 
Bianchi III models, the Kantowski-Sachs models predict
a quadrupolar modulation in the CMB with all the features observed in the data, 
namely the sign of the bipolar coefficients $A_{ll'}^{LM}$, 
their acoustic oscillations, and a particular relation between them. 
To obtain the observed amplitude requires, however, an anisotropic
curvature at a level of $\Omega_{\rm curv} \simeq 10^{-2}$,
which is incompatible with the bound derived from the value of the 
quadrupole, namely $\Omega_{\rm curv} \lesssim 10^{-4}$.
We briefly argued that this bound could be mitigated.
The price to pay is unfortunately an increased complexity of the model, which 
we leave for future work. 
We conclude that the model in its present, most simple, form is 
unable to account for any of the anomalies.

As an additional line for future investigation 
we mention the generation of primordial magnetic fields. 
Since KS spacetimes are not conformally
flat, we indeed expect that long wavelength magnetic fields are
produced from vacuum fluctuations  
during inflation, without recourse to noncanonical couplings of masses.
This generation could perhaps be sufficient to explain the primordial 
magnetic fields necessary
to seed galactic dynamos.


\acknowledgments

We thank Michael Salem and Matthew Johnson for valuable comments on 
the first preprint of this article.
JCN would also like to thank Raphael Bousso, Ben Freivogel, Roni Harnik, Matthew Johnson,
Antony Lewis, and I-Sheng Yang for helpful discussions, and
acknowledges the hospitality of the Berkeley Center for Theoretical
Physics and the Aspen Center for Physics where many of them took place.
We further wish to thank 
the anonymous referee for useful comments and 
Aleksandar Raki\'c for some insights on the CMB anomalies.
The work of JA was supported by the German Research Foundation (DFG) through
the Research Training Group 1147 ``Theoretical Astrophysics and
Particle Physics''. 

\begin{appendix}

\section{Kantowski-Sachs spacetimes}
\label{app:KS}

\subsection{Geometry}

We collect some useful properties of Kantowski-Sachs (KS) spacetimes \cite{KS}.
They are (locally) homogeneous, that is they are invariant under an
isometry group $G$ which acts transitively on the spacelike hypersurfaces
(i.e.\ for two points $x$ and $y$, there exists an isometry $g \in G$ such that 
$g \circ x = y$). The group has four parameters and admits a three-parameter
subgroup isomorphic to $SO(3)$. 
The Killing vectors therefore verify the Lie algebra
$\left[ X_i, \, X_j \right] = \epsilon_{ijk} X_k$ with $i=1,2,3$
and $\left[ X_4,\, X_i \right] = 0$.
Moreover there exists locally a coordinate system $(t,z,\theta,\phi)$ 
which diagonalizes the metric,
\begin{equation} \label{metricKS}
  ds^2 = - dt^2 + a^2(t) dz^2 + b^2(t) \left( d\theta^2 + \sin^2\theta\, d\phi^2  \right) ~,
\end{equation}
where $z$ is the coordinate of the flat direction, and $(\theta,\phi)$ are 
the intrinsic coordinates on a two-sphere.
Then $X_4 = \partial/\partial z$ and $X_i$ are the familiar generators of the rotation group.
Because of the spherical symmetry, and since it is not conformaly flat, 
KS is of Petrov type D. The principal null directions are in the plane $(t,z)$.

It is good to make a pause here and compare with the perhaps better known
Bianchi classes. Bianchi class I, V, VII and IX have a universal covering with 
constant spatial curvature. Bianchi class I is probably the most studied
anisotropic space to date because it has locally Euclidean 
spatial sections \cite{DulaneyGresham,PPU1,PPU2,GCP}.
Classes II, IV and VIII do not have an obvious cosmological application.
Finally class III resembles closest KS spacetimes. The spatial metric is
\begin{equation}
  ds_{\rm B \, III}^2 = -dt^2 + a^2(t) dz^2 + b^2(t) \left( d\chi^2 + \sinh^2\chi \, d\phi^2  \right) ~.
\end{equation}
Like KS spacetime, the orbits are two dimensional.
The only difference seems to be the sign of the curvature of the latters, the topology of the 
spacelike hypersurfaces being $\mathbb{R} \times \mathrm{H}_2$. 
But the metric is deceptive for one can show that the isometry group of Bianchi III 
admits a second three parameter
subgroup whose orbits are three dimensional, which is not the case 
if the two dimensional orbits have a positive curvature (see in particular 
\cite{KSalgebra} for details). In that sense, Bianchi III and KS 
spacetimes form two distinct classes. 
Note finally that Bianchi III spaces can be compactified 
by identifying points on the spacelike surfaces with respect to a discrete 
subgroup of $G$, 
yielding a multiconnected space, for instance a torus \cite{Fagundes}.

\subsection{Evolution}

Returning to KS spacetimes, the spatial sections are described by the 
Ricci
scalar ${}^3 \mathcal{R} = 1/b^2$ and 
the intrinsic curvature tensor is 
$K^{i}_{\,\,j} = \frac{1}{2} h^{ik} \partial_t h_{kj} = 
{\rm diag}\left( \alpha, \beta, \beta \sin^2\theta  \right)$, with $\alpha = \dot a/a$ 
and $\beta = \dot b / b$.
Its trace, the local expansion rate in the frame (\ref{metricKS}), 
is $K = \gamma^{ij} K_{ij} = \alpha + 2 \beta$. 
The shear, $\sigma^{i}_{\,\,j} = K^{i}_{\,\,j}-\delta^{i}_{\,\, j} K/3 = 
\frac{1}{3}{\rm diag}\left( - 2\delta, \delta, \delta  \right)$ with $\delta = \beta - \alpha$,
characterizes the anisotropy of the expansion.
Einstein's equations with a perfect fluid are 
\begin{eqnarray}
  &&2\alpha \beta + \beta^2 + \frac{1}{b^2} = T_{t t} = \rho \\
  &&2 \frac{\ddot b}{b} + \beta^2 + \frac{1}{b^2} = - T_{z z} = - p_z \\
  &&\frac{\ddot a}{a} + \frac{\ddot b}{b} + \alpha \beta = - T_{\theta \theta} = - T_{\phi \phi} = -p_\theta
\end{eqnarray}
and the Bianchi identities are
\begin{eqnarray} \label{Bianchy id}
  \dot \rho + K \rho + \alpha p_z + 2 \beta p_\theta = 0 \, .
\end{eqnarray}
We can combine the first three equations to obtain equations for the local expansion rate 
and shear 
\begin{eqnarray}
\label{eq1}
  &&\frac{1}{3} \left( K^2 - \delta^2 \right) +  \frac{1}{b^2} = \rho \\ 
\label{eq2}
 &&\dot K + \frac{K^2}{3} + \frac{2}{3} \delta^2 = - \frac{1}{2} (\rho + P) \\
\label{eq3}
  &&\dot \delta + K \delta + \frac{1}{b^2} = 0 
\end{eqnarray}
where we note $P = p_z + 2 p_\theta$.

The autonomous system formed by these equations has been analyzed in 
\cite{KSsing} and the possible singularities were described.
Assuming the equation of state $p_i \simeq - \rho$, 
the case relevant for our inflationary model is a pancake singularity, i.e.\ 
$a \to 0$ and $b \to {\rm cte}$ for $t \to - \infty$, where only 
the non compact direction grows in the first stage dominated by curvature. 
The physical variables behave as ${}^3\mathcal{R} \sim K^2/3 \sim \delta^2/3 \sim \rho$.
Once inflation starts, the curvature term $1/b^2$ in equation (\ref{eq1})
becomes negligible and the universe becomes effectively isotropic and 
asymptotes to de Sitter space, $a \sim b \sim e^{Ht}$. 
To see this, use (\ref{eq1}) to eliminate $\delta^2$ from (\ref{eq2}) and write
$K = 3\dot\Omega/\Omega$ to get $\ddot \Omega / \Omega = 2 \rho /3 \simeq {\rm cte}$.
Hence, $K \simeq {\rm cte} = 3H$ and $\Omega \simeq \Omega_0 \cosh \left( H (t-t_0)\right)$. 
Finally substitution of this solution in (\ref{eq3}) or (\ref{eq1}) gives  
$\delta \propto e^{-2Ht}$.
In view of this solution, it is reasonable to expect that inflation
is an attractor solution as in isotropic cosmologies.

The asymptotic behavior translates as follows in the coordinate system (\ref{RNdSmetric}).
For large enough $R$, $f(R) \simeq 1 - \Lambda R^2 / 3$ becomes a good approximation  and
(\ref{RNdSmetric}) 
coincides with the static de Sitter metric
\begin{multline}
\label{dSmetric}
\dd s^2 = -\left(1 - H^2 R^2\right) \dd t_{\rm s}^2 + 
 \left(1 - H^2 R^2\right)^{-1} \dd R^2\\
+ R^2 \left[\dd \theta^2 + \sin^2 \theta \dd \phi^2\right]~,
\end{multline}
de Sitter space indeed admits a foliation $\mathbb{R} \times \mathrm{S}_2$ obtained 
simply by extending (\ref{dSmetric}) 
to the region where $R$ is timelike, $R > R_c \equiv H^{-1}$,
followed by an obvious change of coordinates which puts then the metric in the form
\begin{equation}
  \dd s^2 = - \dd t^2 + \frac{\sinh^2 Ht}{H^2} \dd z^2  + \frac{\cosh^2 Ht}{H^2}\left(\dd \theta^2 + \sin^2 \theta \dd \phi^2\right) \, .
\end{equation}
Cosmological models based on de~Sitter space are usually studied in 
foliations of a Friedman-Robertson-Walker type, i.e.\ which are both homogeneous and isotropic,
either closed, open, or flat.
This choice is not available in the spacetime of
black hole pairs because symmetry dictates that homogeneous fields 
are functions of $R$, and therefore
homogeneous slices are slices of constant $R$.

From the previous results, we can assume that reheating is an isotropic process, and therefore
the CMB is such that $p_z = p_\theta = \rho/3$. The equation of conservation of the energy
(\ref{Bianchy id}) is then solved trivially $\rho \propto \Omega^{-4}$. From Stefan's law
we get that a surface of constant temperature is
\begin{equation}
  T = {\rm cte} \quad \Leftrightarrow \quad \Omega = {\rm cte}  \quad \Leftrightarrow  \quad t= {\rm cte} 
\end{equation}

Since $\delta \ll 1$, eqs. (\ref{eq1}) and (\ref{eq2}) with the 
terms $\delta^2$ neglected are the familiar
Friedman equations.
For an equation of state $p=(\gamma - 1)\rho$ we thus get in the first approximation 
$a \sim b \sim t^{2/3\gamma}$.
The general solution of equation (\ref{eq3}) is 
\begin{eqnarray} \label{sol delta}
  \delta &=& \delta_0 \left( \frac{\Omega_0}{\Omega}\right)^3 
  + \frac{1}{ab^2}\int_{t_0}^{t}\!\!dt'a(t') 
\nonumber \\
  &\sim& \frac{3\gamma}{3\gamma+2}t^{1- 4/3\gamma}~.
\end{eqnarray}
The second line is obtained after substituting the solutions of $a$ and $b$
previously found and noticing that the 
homogeneous solution is subleading. We could then use (\ref{sol delta}) to iterate an
asymptotic expansion for $a$, $b$ and $\delta$.
Thus during matter domination, the ratio of the shear to the isotropic expansion rate
grows like
\begin{equation} \label{delta/K}
  \frac{\delta}{H} \sim t^{2/3} \sim \Omega~.
\end{equation}
This can be reexpressed in terms of the curvature parameter defined by 
\begin{equation}
  \Omega_{\rm curv} = \frac{1}{b^2 H^2} \sim t^{2/3}~.
\end{equation}
The relation with the parameter $r$ 
of eq.\ (\ref{important}) follows from their respective definitions:
$r= l_{\rm LSS}/R$
is the radius of the last scattering surface 
in the units of the radius of ${\mathrm{S}_2}$, and 
$\Omega_{\rm curv} = (l_{H_0}/R)^2$
is the square of the Hubble length
in those units.
The relation (\ref{important}) follows from the proportionality of 
$l_{\rm LSS}$ and $l_{H_0}$.

We note finally that these results are readily adapted to the Bianchi type III 
since only the sign of the curvature term $1/b^2$ changes.
By contrast in Bianchi I, where the spatial hypersurfaces are Euclidean, 
only the first term on the r.h.s.\ of (\ref{sol delta}) is present, and 
(\ref{delta/K}) is replaced by $\delta/K \sim \Omega^{-3/2}$.

\end{appendix}

\bibliographystyle{h-physrev}
\bibliography{decompact}

\end{document}